%% file: vdwarxiv.tex
\documentclass[aps,pra,groupedaddress,floatfix,twocolumn,showpacs]{revtex4}
\usepackage{bm}
\usepackage{graphicx}
\usepackage{amsmath} 
\usepackage{longtable}

\def\openone{\leavevmode\hbox{\small1\kern-3.3pt\normalsize1}}

\def \ket#1{{\left|#1\right\rangle}}
\def \bra#1{{\left\langle#1\right|}}

\def \bbra#1{{\left\{#1\right|}}

\def \expect#1{{\left \langle #1 \right\rangle}}
\def \overlap#1#2{{\langle#1|#2\rangle}}

\def \CG#1#2#3{{C_{#1\,#2}^{#3}}}

\def\tp#1#2#3{{\left\{#1\otimes#2\right\}_{#3}}}

\def\ninej#1#2#3#4#5#6#7#8#9{{\left\{\begin{array}{@{}c@{}c@{}c@{}} #1&#2&#3\\#4&#5&#6\\#7&#8&#9\end{array}\right\}}}

\def\Varsh#1{{\cite{Varshalovich} (#1)}}

\def\forster{F\"{o}rster}
\def\half#1{{#1\over 2}}
\def\rabi{{\Omega}}

\newcommand \bse{\begin{subequations}}
\newcommand \ese{\end{subequations}}
\newcommand \bea{\begin{eqnarray}}
\newcommand \eea{\end{eqnarray}}

\def\bshift{{\sf B}}
\def\fshift{{\sf  D}}
\def\eref#1{{(\ref{#1})}}
\newcommand\rmat[2]{{R _{#1}^{#2}}}

\begin{document}

 \title{Consequences of Zeeman Degeneracy for van der Waals \\ Blockade between Rydberg Atoms}
 
 \author{Thad G. Walker and M. Saffman}
 
 \affiliation{Department of Physics, University of Wisconsin-Madison, Madison, Wisconsin 53706}

\begin{abstract}We analyze the effects of Zeeman degeneracies on the long-range interactions between like Rydberg atoms, with particular emphasis on applications to quantum information processing using van der Waals blockade.  We present a general analysis of how degeneracies affect the primary error sources in blockade experiments, emphasizing that blockade errors are sensitive primarily to the weakest possible atom-atom interactions between the degenerate states, not the mean interaction strength.  We present explicit calculations of the van der Waals potentials in the limit where the fine-structure interaction is large compared to the atom-atom interactions.  The results are presented for all potential angular momentum channels invoving s, p, and d states.  For most channels there are one or more combinations of Zeeman levels that have extremely small dipole-dipole interactions and are therefore poor candidates for effective blockade experiments.  Channels with promising properties are identified and discussed.  We also present numerical calculations of Rb and Cs dipole matrix elements and relevant energy levels using quantum defect theory, allowing for convenient quantitative estimates of the van der Waals interactions to be made for principal quantum numbers up to 100.  Finally, we combine the blockade and van der Waals results to quantitatively analyze the angular distribution of the blockade shift and its consequence for angular momentum channels and geometries of particular interest for blockade experiments with Rb.
\end{abstract}

\date{\today}
  
 \maketitle
 
 \section{Introduction}
 
 The interactions between   ultracold Rydberg atoms have  been of interest for some time now, beginning experimentally with studies of resonant energy transfer \cite{Anderson98,Mourachko98}.
 The strong, long-range interactions between Rydberg atoms are recognized as being extremely important for understanding these phenomena, such as the evolution of clouds of ultracold Rydberg atoms into ultracold plasmas and vice  versa \cite{Robinson00,Killian01}.  Compelling theoretical concepts for exploiting  Rydberg-Rydberg interactions for quantum information processing using single-atom qubits\cite{Jaksch00} and atomic ensembles \cite{Lukin01} have stimulated further experimental \cite{Farooqi03,Afrousheh04,Singer04,Tong04,Deiglmayr05,Li05,Liebisch05,Singer05b,Afrousheh06,Amthor07,Bohlouli07,Carroll06,Heidemann07,Overstreet07,Vogt07,Johnson07} and theoretical \cite{Saffman02,Saffman04,Saffman05b,Singer05,Walker05,Brion07} work.
 
 One of the key ideas behind the potential for coherent quantum information applications using Rydberg atoms is the concept of dipole blockade\cite{Lukin01}.  If an ensemble of $N$ ground state atoms is resonantly driven by a laser tuned to a Rydberg state, the excitation proceeds first through states with a single atom excited, then through states with two atoms excited, and so on.  If  the Rydberg-Rydberg interaction energies are sufficiently strong, excitation of the multiply-excited atom states will be greatly suppressed, and only a single atom will be excited at a time.  Excitation of multiple atoms is blocked by the dipole-dipole interactions between the Rydberg atoms.  In this blockade regime, an effective two-level system is realized between the ground state $\ket{ggg}$ (for $N=3$) and the entangled symmetric excited state $\ket{\sf s}=(\ket{gge}+\ket{geg}+\ket{egg})/\sqrt{3}$.  This two-level system has an effective light-atom coupling that is a factor of  $\sqrt{N}$ larger than the light-single-atom coupling, and is promising for a wide variety of quantum manipulation applications.
 
\begin{figure}[t] 
   \centering
   \includegraphics[width=2.5in]{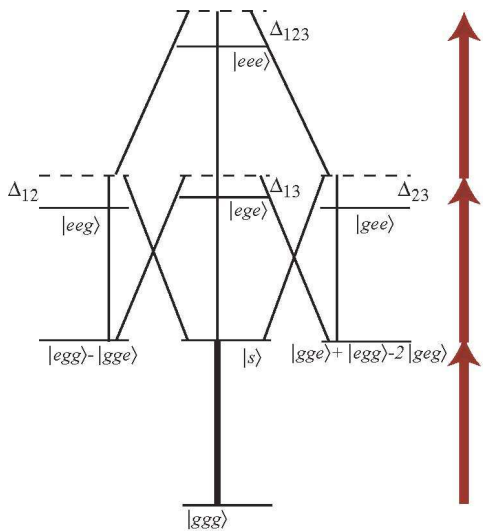} 
   \caption{(color online) Energy levels of 3 atoms arranged in a line, with dipole-dipole interactions $\Delta_{12}=\Delta_{23}>\Delta_{13}$. The dashed lines are the energy level positions if the dipole-dipole interactions were zero, and the arrows denote the successive excitation of the multiple atom states by the resonantly-tuned laser.  Solid lines indicate pairs of states that are coupled by allowed electric dipole transitions.  All these transitions are off-resonant save the $\ket{ggg}-\ket{s}$ pair that consititute an effective two-level system in the presence of blockade.}
   \label{eleveldiagram}
\end{figure}

Some of the issues associated with realizing dipole blockade are illustrated in the energy level diagram for the case $N=3$ in Figure~\ref{eleveldiagram}.  The atoms are assumed to be placed symmetrically along a line so the dipole-dipole interactions between atom pairs  are $\Delta_{12}=\Delta_{23}>\Delta_{13}$.  
 Blockade is effective if and only if each of the dipole-dipole interaction strengths is much greater than the atom-light coupling.  In that case, all the two- and three- atom excited states are out of resonance with the laser and an effective two-level system results.  The influence of the two-atom states is limited to a small AC-Stark shift of the resonance frequency.  However, if $\Delta_{13}$ is insufficiently large then the doubly-excited state $\ket{ege}$ becomes appreciably populated.  In addition,  $\ket{ege}$ is also resonantly coupled to other singly-excited states  that do not couple optically to the ground state.  Thus the primary errors that enter the blockade process are the production of doubly-excited states \cite{Lukin01} and singly-excited states outside the desired two-level system.
 
For the case of non-degenerate Rydberg states, these errors have been discussed before\cite{Lukin01,Saffman02}.  In Section~\ref{block} of this paper, we extend the analysis to the more realistic case of degenerate alkali Rydberg states interacting via  van der Waals forces.  The types of errors are qualitatively the same, but we show explicitly how to account for degeneracies in the Rydberg energy levels.  We show that the figure of merit for blockade is primarily determined by the {\it weakest} of the many potential curves that emanate from a given degenerate Rydberg state.  In many cases there are one or more of these degenerate states with nearly zero van der Waals interaction and hence weak blockade, despite the state-averaged van der Waals interaction being substantial.  These  ``\forster-zero'' states \cite{Walker05} can be thought of as resulting classically from precession of the atomic angular momenta due to the dipole-dipole interaction.  These states
did not appear in Ref.~\cite{Reinhard07} due to the use of non-degenerate perturbation theory, effectively ignoring the ability of the dipole-dipole interaction to change the orientation of the individual atomic angular momenta.

With an eye toward applying these ideas to Rydberg states in the $30<n<100$ range of interest for quantum manipulation of atoms at optically resolvable distances, we are led in Section~\ref{vdWsection} to consider van der Waals interactions at distances where they are weak compared to the fine-structure splitting, a limit rarely treated in the literature\cite{Walker05,Stanojevic06, Reinhard07}.  We calculate the van der Waals eigenvalues for 21 different angular momentum channels likely to be encountered for low-angular momentum Rydberg states, and present the numerical values of matrix elements and energy defects so that quantitative estimates of blockade can be made by others.
  We follow this in Section \ref{sec:angle} with two concrete calculations of the blockade frequency shifts to be expected for the case of a linear geometry, and the angular distributions for interesting cases.

  We note that in this paper we are using the term ``blockade'' in a more restrictive sense than has been used in a number of experiments to date.  It has become common to refer to a suppression or reduction in the number of excited Rydberg atoms due to atom-atom interactions as blockade.  However, such effects do not necessarily imply the type of blockade necessary to produce collective singly-excited quantum states that are of primary interest for this paper.

\section{Errors in Quantum Manipulations Using Blockade}\label{block}

\def\calV{{\mathcal V}}

In our previous work \cite{Saffman02} on single atom and single photon sources, we assumed spatially uniform Rabi frequencies and a
spherical distribution of atoms. We also, in order to keep things simple, assumed the effective Rydberg shift for each atom was the
same, and assumed a non-degenerate Rydberg state. We wish to remove these assumptions in order to more accurately describe  situations that will be faced in Rydberg blockade experiments.

We will assume that the blockade shifts are large enough to limit a cloud of $N$ atoms to at most 2 excited atoms.  Thus the atomic cloud can be in the possible states
\begin{equation}
\ket{g},\ket{\gamma k},\;{\rm  and }\; \ket{\varphi kl}
\end{equation}
representing respectively all the atoms in the ground state, the $k$th atom in the singly-excited Rydberg state $\gamma$, and the $k$th and $l$th atoms in the doubly-excited Rydberg state $\varphi$.

We assume that the coupling of the excitation light to the atoms can be represented by an effective excitation operator 
\begin{equation}
A^{\dagger}=\half1\sum_{k}\calV_{k}^{\dagger}\end{equation}
 that takes atoms at various positions $k$ from the ground state to a Zeeman sublevel $\ket{\gamma{k}}$ of the Rydberg state, and it also takes states with one Rydberg atom excited at position $k$ to a doubly-excited state.  The Rabi coupling to the state $\ket{\gamma k}$ is 
  \begin{equation}
\Omega_{\gamma k}=\bra{\gamma k}\calV_{k}\ket{g}.
\end{equation}
Here and in the remainder of the paper we use units with $\hbar=1$.

Generalizing Saffman and Walker\cite{Saffman02}, the wavefunction for the $N$-atom ensemble is
\begin{equation}
|\psi\rangle=c_{g}|g\rangle+\sum_{rk}c'_{\gamma k}|\gamma{k}\rangle+\sum_{\varphi,k<l}c_{\varphi kl}\ket{\varphi{kl}}
\end{equation}
where the doubly-excited states $\ket{\varphi{kl}}$ are eigenstates of the effective Rydberg-Rydberg Hamiltonian:
\begin{equation}
H_{\rm eff}\ket{\varphi{kl}}=\Delta_{\varphi kl}\ket{\varphi{kl}}
\end{equation}
The index $\varphi$ represents the various possible Rydberg-Rydberg shifts $\Delta_{\varphi kl}$ for atoms  $k$ and $l$ in the sample.  For the specific situations treated in this paper, $H_{\rm eff}$ represents van der Waals interactions and so rotational invariance guarantees that $\Delta_{\varphi kl}$ depends only on the index $\varphi$ and the distance $R$ between atoms $k$ and $l$.  Likewise, the states $\ket{\varphi{kl}}$ for different atom pairs are independent of $k$ and $l$ when expressed in a coordinate system oriented along the interatomic axis; in the laboratory frame they are related to each other by rotations.  We will not use these properties in the remainder of this section, so the results also apply to cases where the effective Rydberg-Rydberg interactions are modified by external laboratory fields that do not preserve their rotational invariance.

It is convenient to introduce the wavefunction for a symmetric singly-excited state as follows:
\begin{equation}
\ket{\sf s}={1\over \Omega_{N}}\sum_{k}\calV^{\dagger}_{k}\ket{g}=\sum_{\gamma k}{\rabi_{\gamma k}\over \rabi_{N}}\ket{\gamma k}
\end{equation}
and orthogonalize the remaining singly-excited states as
\begin{equation}
\ket{(\gamma k)_\perp}=a_{\gamma k}\left(\ket{\gamma k}-\overlap{{\sf s}}{\gamma k}\ket{\sf s}\right)
\end{equation}
where $a_{\gamma k}$ is a normalization factor.  These orthogonalized states do not couple directly to $\ket{g}$ via the light, but are populated only by coupling through the doubly-excited states. The collective Rabi frequency is defined to be
\begin{equation}
\rabi_N=\sqrt{\sum_{\gamma k}\left| \rabi_{\gamma k}\right|^2}=\sqrt{N}\Omega_{0}
\end{equation}
where $\Omega_{0}$ is the rms single-atom Rabi frequency averaged over the Rabi frequencies of the individual atoms.

The symmetric singly-excited state is coupled by the light to the ground state and the doubly-excited states.  The matrix elements are
\begin{eqnarray}
\bra{\sf s}A^{\dagger}\ket{g}&=&\rabi_{N}/2\\
\bra{\varphi kl}A^{\dagger}\ket{\sf s}&=&{1\over 2\Omega_{N}}\sum_{k'l'}\bra{\varphi kl}\calV^{\dagger}_{k'}\calV^{\dagger}_{l'}\ket{g}\nonumber \\
&=&{\bra{\varphi kl}\calV^{\dagger}_{k}\calV^{\dagger}_{l}\ket{g}\over \rabi_{N}}\equiv{\rabi_{N}\over N}\kappa_{\varphi kl}
 \end{eqnarray}
where the factor of 2 disappeared because both $k'=k,l'=l$ and $k'=l,l'=k$ terms contribute to the sum.  The dimensionless overlap factor $\kappa_{\varphi kl}$ gives the relative amplitude for exciting a particular  pair $kl$ of atoms to the doubly-excited Rydberg state $\varphi$.  It depends on the experimental geometry (laser polarization, spatial variation of intensity, etc.) and in particular depends on the relative orientation of the atom pair $kl$ and the light polarization.

Using the above definitions, the Schr\"odinger equations for the ground, symmetric first-excited state, and the second excited states are
\begin{eqnarray}
i{\dot{c}}_{g} & = &\bra{g}A\ket{\sf s}={\rabi_{N}\over 2} c_{\sf s} \\
i{\dot{c}}_{\sf s} & = & \bra{\sf s}A^{\dagger}\ket{g}c_{g}+\sum_{\varphi kl}\bra{\sf s}A\ket{\varphi kl}c_{\varphi kl}\nonumber \\
&=&{\rabi_{N}\over 2} c_{g}+\sum_{\varphi kl} {\rabi_{N}\over N}\kappa^{*}_{\varphi kl}c_{\varphi kl}\label{se1}\\
i{\dot{c}}_{\varphi kl} & = & \Delta_{\varphi kl}c_{\varphi kl}+\bra{\varphi kl}A^{\dagger}\ket{\sf s} c_{\sf s}\nonumber \\
&=& \Delta_{\varphi kl}c_{\varphi kl}+{\rabi_{N}\over N}\kappa_{\varphi kl}c_{\sf s}
 \label{se2}
\end{eqnarray}
where we have omitted the possible excitation of the doubly-excited states from the orthogonalized singly-excited states.

We solve these equations as a successive approximation in the ratios $\rabi/\Delta$.  The doubly-excited amplitudes are of
order $\rabi/\Delta$ so to zeroth approximation we take $c_{\varphi kl}=0$ (perfect blockade approximation) and get
\begin{equation}
i{\dot{c}}_{g}={\Omega_{N}\over 2}c_{\sf s}; \;\;\; i\dot{c_{\sf s}}={\Omega_{N}\over 2}c_{g}
\end{equation}
which are the equations for standard Rabi flopping at collective Rabi frequency $\Omega_{N}$.  Explicitly,
 a resonant Rabi pulse of duration $T$ beginning with $c_g=1$, has the solution 
\begin{eqnarray}
{c}_{g} & = &\cos( {\rabi_N T/2})\\
c_{\sf s} & = & -i\sin(\rabi_N T/2)\label{se1sol}
\end{eqnarray}  In the approximation of perfect blockade, the system undergoes collective Rabi flopping without dephasing or loss of population to other states.

Let us now consider the effectiveness of Rydberg blockade.  We start by calculating the probability of excitation of more than one atom.  In the  limit of large but finite Rydberg-Rydberg shift, we can make an adiabatic approximation to Eq.~(\ref{se2}) to get
\begin{equation}
c_{\varphi kl}=-{\rabi_{N}\kappa_{\varphi kl}\over N\Delta_{\varphi kl}}c_{\sf s}
\label{adiab}
\end{equation}
The probability of double excitation is
\begin{equation}
P_2=\sum_{\varphi;k<l}\left|c_{\varphi kl}\right|^2={\rabi_N^{2}\over N^{2}}\sum_{\varphi;k<l}\left|{\kappa_{\varphi kl}\over \Delta_{\varphi kl}}\right|^2
\end{equation}
It is critical to note that, given relatively even excitation of the two-atom Rydberg states,  it is an average of $1/\Delta_{\varphi kl}^2$ that determines the blockade
effectiveness.  This means that Rydberg-Rydberg states with small van der Waals shifts are much more strongly weighted than those with large energy shifts.  Let us define a mean blockade shift $\bshift$ via
\begin{equation}
{1\over \bshift^2}={2\over
N(N-1)}\sum_{\varphi;k<l}{\kappa_{\varphi kl}^{2}\over\Delta_{\varphi kl}^2}
\label{blockshift}
\end{equation}
Then the probability of double excitation becomes
\begin{equation}
P_2={(N-1)\rabi_N^2\over  2N\bshift^2}
\label{pdouble}
\end{equation}
This shows that, for fixed $\rabi_N$, the probability of double excitation is virtually independent of the number of atoms in the
ensemble.  This does not contradict Fig. 3 of Ref.~\cite{Saffman02}, where the plot assumed a fixed value of the single-atom Rabi frequency, not
$\rabi_N$.  We will evaluate $\bshift$ for cases of experimental interest below.

It is  is important to keep in mind that the blockade shift $\bshift$ depends on the polarization of the excitation light as well as the Zeeman structure of the state $\ket{g}$ 
through the overlap factor $\kappa_{\varphi kl}$.  We do not explicitly indicate these dependences in order to avoid a proliferation of subscripts.  Explicit examples will be given in Section~\ref{sec:angle}.

In addition to the production of population of doubly-excited states, finite blockade also causes a frequency shift of the effective two-level system through virtual excitation of the doubly-excited states.  Using the adiabatic approximation results for $c_{\varphi kl}$ modifies Eq.~\eref{se1} to  
\begin{equation}
i{\dot{c}}_{\sf s}={\rabi_{N}\over 2} c_{g}-{\rabi_{N}^{2}\over N^{2}}\sum_{\varphi; k<l} {|\kappa_{\varphi kl}|^{2}\over \Delta_{\varphi kl}} c_{\sf s}\label{se1p}
\end{equation}
The second term represents a shift in the resonance frequency of the effective two-level system.  Defining a frequency shift factor
\begin{equation}
{1\over \fshift}={2\over N(N-1)}\sum_{\varphi; k<l} {|\kappa_{\varphi kl}|^{2}\over \Delta_{\varphi kl}}
\end{equation}
the resonance frequency of the two-level system is shifted by 
\begin{equation}
\delta\nu={(N-1)\Omega_{N}^{2}\over 2N \fshift}
\end{equation}

At this level of approximation, the final blockade error that can result is to transfer population out of the ``computational basis''  of $\ket{g}$ and $\ket{\sf s}$ and into the other singly-excited states $\ket{(\gamma k)_\perp}$.  The amplitude for these states obeys
\begin{eqnarray}
i \dot c_{\gamma k\perp}&=&\sum_{\varphi k'l'}\bra{(\gamma k)_\perp}A\ket{\varphi k'l'}c_{\varphi k'l'}\\
&=& -{\rabi_N\over N}\sum_{\gamma l}\bra{(\gamma k)_\perp}{\cal V}_l\ket{\varphi kl}{\kappa_{\varphi kl}\over \Delta_{\varphi kl}}c_{\sf s}
\end{eqnarray}
in the adiabatic approximation (\ref{adiab}).  This does not simplify as nicely as Eq.~(\ref{blockshift}), but a good estimate of the probability of excitation of the singly-excited states outside the computational basis is
\begin{equation}
P'_1\sim P_2
\end{equation}
Thus in estimating the blockade errors one should rough\-ly double the estimate (\ref{pdouble}) obtained from the blockade shift and the collectively enhanced Rabi frequency.

 \section{van der Waals Interactions Between Degenerate Rydberg Atoms}\label{vdWsection}

 \subsection{General Discussion}
\label{general}
 
 The electrostatic dipole-dipole interaction between two non-overlapping atoms $A$ and $B$ that lie a distance $R$ apart along the z-axis is
 \begin{equation}
V_{\rm dd}={e^{2}\over R^{3}}\left(\bm a\cdot\bm b-3a_zb_z\right)={-\sqrt{6}e^{2}\over R^{3}}\sum_{p}\CG{1p}{1\bar p}{20}a_{p}b_{\bar p}
\end{equation}
where $\bm a$ is the position of the electron on atom $A$ and $\bm b$ is the position of the electron on atom $B$, and $\bar p=-p$. The odd parity of the dipole operators $\bm a$ and $\bm b$ results in an initial 2-atom state $\ket{n_{A}l_{A}j_{A}m_{A}n_{B}l_{B}j_{B}m_{B}}$ being mixed with states of $l_{A}\pm 1$ and $l_{B}\pm 1$.  The individual total angular momenta $j_A$ and $j_B$ may also change, consistent with dipole selection rules, as do the principal quantum numbers $n_A$ and $n_B$. The total projection of the angular momentum along the z-axis $M=m_{A}+m_{B}$ is conserved, but the individual quantum numbers change by $\pm 1$ or 0.

For the simplest version of blockade physics, we are interested in the case where the two atoms are being excited to the same energy level, so that $n_{A}=n_{B}=n$, $l_{A}=l_{B}=l$, $j_{A}=j_{B}=j$.  Then the dipole-dipole interaction causes the reaction
\begin{equation}
nl{j}+nl{j}\rightarrow n_sl_sj_s+n_tl_{t}j_t
\end{equation}
with an energy difference between the final ($s,t$) and initial two-atom states  
\begin{equation}
\delta=E(n_{s} l_{s}j_{s})+E(n_{t}l_{t}{j_{t}})-2E(nl{j})\end{equation}
 that we will call a \forster\ defect.
At the largest distances, a non-zero \forster\ defect causes  the dominant long-range interaction between the atoms to be of the $R^{-6}$ van der Waals type that arises from $V_{\rm dd}$ in second order.  (The $R^{-5}$ quadrupole-quadrupole interaction is normally much smaller, as estimated in   Appendix \ref{quadquad}.)  For atoms that are closer, the van der Waals interaction becomes large enough to mix the fine-structure levels together, particularly for the d levels.  This occurs (for 30--80d levels) in the 0.8--8 $\mu$m range of distances of interest for interactions between optically resolvable Rydberg atoms.  For atoms at somewhat smaller distances, typically 0.5--5 $\mu$m,  the dipole-dipole interaction is comparable to the energy differences between nearby states, so the interactions become resonant  and vary as $R^{-3}$. 

There are some situations where the \forster\ defects are smaller than the fine-structure splitting, a notable example \cite{Bohlouli07} being at 43d in Rb, where the reaction $43d_{5/2}+43d_{5/2}\rightarrow 45p_{3/2}+41 f$ has less than 10 MHz \forster\ defect while the 43d fine-structure splitting is 150 MHz.  In this case the transition to resonant dipole-dipole coupling occurs at longer range than fine-structure  mixing.

As discussed above, the possibility of using Rydberg blockade for mesoscopic quantum manipulation depends critically on the most weakly interacting atoms, those that are furthest apart.  It is therefore extremely important to understand the interactions of Rydberg atoms in the limit where the van der Waals interaction has not mixed the fine-structure.  For most of the rest of this paper we will restrict our discussions to this limit.

For a non-zero \forster\ defect,  the energy shifts of the initial states are determined at long range by the effective second-order perturbation operator or van der Waals interaction
\begin{equation}
H_{{\rm{vdW}}}  = \sum\limits_{st} {\frac{{V_{{\rm{dd}}} \left| {st} \right\rangle \left\langle {st} \right|V_{{\rm{dd}}} }}{{-\delta_{st} }}} 
\label{vdW}
\end{equation}
This operator is understood to act within the degenerate set of Zeeman sublevels of the two-atom initial state.  In general, the sum is over the various intermediate two-atom energy levels $\ket{st}$ that obey the selection rules of $V_{dd}$ discussed above, and $\delta_{st}$ is the \forster\ defect for each channel with respect to the initial state.  The indices $s$ and $t$ denote the full set of quantum numbers that specify the intermediate states.
In practice, this sum can be greatly simplified by noting that in most cases only states close in energy to the initial state have significant $r$ matrix elements and therefore dominate the matrix elements of $V_{\rm dd}$, and that  typically one or two of these states have  the smallest \forster\ defects.  Thus there will usually be only a couple of intermediate states that give by far the biggest contributions to the van der Waals interactions.

It is particularly important, as we will see, to properly account for the Zeeman degeneracy of the initial and intermediate states.  This degeneracy has the consequence that one cannot calculate the energy shifts by simply taking the expectation values of the van der Waals operator of Eq.~\eref{vdW}.  The dipole-dipole interaction couples an initial state with magnetic quantum numbers $m_{A},m_{B}$ to intermediate states $m_{A}+p,m_{B}-p$ which can then couple to a different Zeeman combination $m_{A}+p+q,m_{B}-p-q$ where $p$ and $q$ range from $-1$ to $1$.  Thus the van der Waals interaction, being second order in $V_{dd}$, changes the magnetic quantum numbers of the individual atoms by up to $\pm2$ units.  It is therefore necessary to use degenerate second-order perturbation theory \cite{Schiff} (Section 31) to understand the van der Waals interactions of Rydberg atoms.  The diagonalization of the effective Hamiltonian of Eq.~\eref{vdW} leads to a range of eigenvalues, meaning that the strength of the van der Waals interactions depend strongly on the Zeeman sublevels.

As we have shown previously \cite{Walker05}, many states which are nearly \forster\ resonant have linear combinations of Zeeman sublevels with zero dipole-dipole coupling. 
The existence of these states can be understood  from angular momentum arguments.    If the initial state atoms have individual angular momenta $j$, these can be coupled together to make a total of $2j+1$ possible states of angular momentum $J$ with $M=0$. (Most of the \forster\-zero states have $M=0$ so we restrict our argument to that case.) Similarly, if the dipole-dipole coupled states are constructed from angular momenta $j_{s}$ and $j_{t}$, with $j_{s}\le j_{t}$, there are $2j_{s}+1$ $M=0$ intermediate states.  The condition for a ``\forster\ zero'' state $\psi_{F}$ is
\begin{equation}
\bra{st}V_{\rm dd}\ket{\psi_{F}}=\sum_{J=0}^{2j}\bra{st}V_{\rm dd}\ket{J}\overlap{J}{\psi_{F}}=0
\end{equation}
which must hold for all of the $2j_{s}+1$ coupled states $\bra{st}$.  The \forster-zero condition is therefore a set of $2j_{s}+1$ equations in $2j+1$ unknowns.  A \forster-zero solution exists  for $j_{s}<j$, and does not exist for $j_{s}>j$.  For $j_{s}=j$ there are not always precise zeros but in the cases studied here there is always a state with very small dipole-dipole coupling.

The above arguments strictly hold for cases where a single channel dominates the dipole-dipole interactions.  Additional channels can give some dipole-dipole coupling to the \forster\ zero states, but the states are still very weakly interacting, leading to potential problems for blockade applications.
 
\subsection{van der Waals Interactions with Fine Structure}

 We now proceed to calculation of the dipole-dipole interaction at such long interatomic distances that the atomic fine-structure is not affected by the dipole-dipole interaction. We assume that all the intermediate states that are coupled to the initial states have the same angular momentum structure, {\it i.e.} that a single virtual process
 \begin{equation}
nl{j}+nl{j}\rightarrow n_{s}l_{s}j_{s}+n_{t}l_{t}{j_{t}}
\end{equation}
occurs where various values of $n_s$ and $n_t$ may contribute but only a single value of $l_s,j_s,l_t,j_t$.
 We denote the initial states of the two atoms $\ket{nlj m_{A}nljm_{B}}\equiv\ket{m_{A}m_{B}}$ and the dipole-dipole coupled intermediate states $\ket{n_{s}l_{s}j_{s}m_{s}n_{t}l_{t}j_{t}m_{t}}\equiv\ket{m_sm_t}$.
 
The dipole matrix elements in $V_{\rm dd}$ can be written in terms of radial matrix elements and angular momentum factors using the Wigner-Eckart theorem \Varsh{Eq. 13.1.5(40)}:
\begin{eqnarray}
\left\langle {n'l'j'm'} \right|r_p \left| {nljm} \right\rangle  &=& \left( { - 1} \right)^{j + l' - \half1} C_{jm1p}^{j'm'} \sqrt {2j + 1}\nonumber\\
&& \times\left\{\begin{array}{@{}*{3}{c}@{}}
   l & {\half1} & j  \\
   {j'} & 1 & {l'}  \\
\end{array}
\right\}\bra{n'l'}|r|\ket{nl}\end{eqnarray}
where the reduced matrix element is
\begin{eqnarray}
\bra{n'l'}|r|\ket{nl}&=&\sqrt {2l + 1} C_{l010}^{l'0} \rmat{nl}{n'l'} 
  = \sqrt {2l + 1} C_{l010}^{l'0}\nonumber \\
  &&\times\int {rP_{n'l'} (r)P_{nl} (r)dr} 
\end{eqnarray}
The radial wavefunctions $P_{nl}(r)$ can be calculated numerically using quantum defect theory or model potentials.
It is convenient to define an operator ${\mathcal M}$ that includes all the angular momentum properties of  $V_{dd}$ for the intermediate states of  angular momenta $j_{s}, j_{t}$:
\begin{eqnarray}
\left\langle {m_{s}m_{t}} \right|{\mathcal M} \left| {m_A m_B } \right\rangle  = \left( { - 1} \right)^{2j + 1}C_{l010}^{l_{s}0}C_{l010}^{l_{t}0}\sqrt{6}(2l+1)&& \nonumber \\
\times  (2j + 1)
\left\{ {\begin{array}{@{}*{3}{c}@{}}
   l & \half1 & j  \\
   {j_s } & 1 & {l_s }  \\
\end{array}} \right\}\left\{ {\begin{array}{@{}*{3}{c}@{}}
   l & \half1 & j  \\
   {j_t } & 1 & {l_t }  \\
\end{array}} \right\}&&\nonumber\\
\times\sum_{p} C_{1p1\bar p}^{20} C_{jm_A 1p}^{j_s m_s } C_{jm_B 1\bar p}^{j_t m_t }\;\;&&
\end{eqnarray} (The matrix elements of $\mathcal M$ are much simpler in a coupled basis (Appendix \ref{coupled}) rather than the product basis we are using here, but for connection to blockade physics the product basis is more convenient.)  
This allows the degenerate Hamiltonian due to the van der Waals interaction to be written
\begin{equation}
H_{\rm vdW}={C_{6}\over R^{6}}\sum_{m_{s}m_{t}}{\mathcal M}^{\dagger}\ket{m_{s}m_{t}}\bra{m_{s}m_{t}}{\mathcal M}={C_{6}\over R^{6}}{\mathcal D}\label{heff2}
\end{equation}
where $C_{6}$  depends only on the atomic energy level structure and radial matrix elements:
\begin{equation}
C_{6}=\sum_{n_{s}n_{t}}{e^{4}\over -\delta_{st}}\left(\rmat {nl}{n_{s}l_{s}}\rmat{nl}{n_{t}l_{t}}\right)^{2}
\label{overallc6}
\end{equation}
The operator ${\mathcal D}=\mathcal M^{\dagger}\mathcal M$ contains all the angular momentum properties of the states.  Its $(2j+1)^{2}$ eigenvalues $D_{\varphi}$, when multiplied by $C_{6}$, give the long-range energies of the two-atom eigenstates:
\begin{equation}
H_{\rm vdW}\ket{\varphi}={C_{6}\over R^{6}}D_{\varphi}\ket{\varphi}
\label{eigen}
\end{equation}
 The eigenvalues $ D_{\varphi}$ obey $0\le D_{\varphi}< 1$.  The sign of the \forster\ defects determines the sign of $C_{6}$.
  In the case that channels of different  angular momentum structure contribute significantly to the long-range interactions, the $H_{\rm vdW}$ matrices for each channel should be computed separately, added together, and then diagonalized.

 We have calculated the eigenvalues (Table \ref{evals}) and eigenvectors \cite{epaps} of $\mathcal D$ for initial $s$, $p$, and $d$ states with fine structure, corresponding to 23 different angular momentum channels.  We will now discuss a few interesting cases before proceeding to using these results for blockade estimates.

 \def\reaction#1#2#3#4#5#6{{$\begin{array}{l}#1_{#2}+#1_{#2}\rightarrow\\\multicolumn{1}{r}{#3_{#4}+#5_{#6}}\end{array}$}}
\begin{table*}[p]
\caption{Relative interaction strengths for van der Waals interactions of Rydberg atoms, for various collision channels.   The potential energy at distance $R$ is the product  $C_6D_{\varphi}/R^{6}$, which contains the effects of Zeeman degeneracy, with the overall $C_{6}$ coefficient (Eq.~\protect\ref{overallc6})  for a particular channel that depends only on the energy level structure and radial matrix elements.  Cases where the $j$ quantum number is not included in the channel description are the sum over fine-structure components of the final state.}\label{evals}
\begin{tabular}{|c|l|c|ll|}\hline
Channel&$|M|$ $\{D_{\varphi}\}$&Channel&$|M|$ $\{D_{\varphi}\}$&\\\hline
\reaction{s}{1/2}{p}{}{p}{}&$\begin{array}{ll}
1 & \{1.33\} \\
 0 & \{1.33, 1.33\}
\end{array}$&
\reaction{s}{1/2}{p}{1/2}{p}{1/2}&$\begin{array}{ll}
 1 & \{0.0988\} \\
 0 & \{0.395, 0\}
\end{array}$&\\\hline
\reaction{s}{1/2}{p}{1/2}{p}{3/2}&$\begin{array}{ll}
 1 & \{0.346\} \\
 0 & \{0.444, 0.0494\}
\end{array}$&
\reaction{s}{1/2}{p}{3/2}{p}{3/2}&$\begin{array}{ll}
 1 & \{0.543\} \\
 0 & \{0.84, 0.444\}
\end{array}$&\\ \hline\hline
\reaction{p}{1/2}{s}{1/2}{s}{1/2}&$\begin{array}{ll}
1 & \{0.0988\} \\
 0 & \{0.395, 0\}\end{array}$&
\reaction{p}{1/2}{s}{1/2}{d}{3/2}&$\begin{array}{ll}
 1 & \{0.346\} \\
 0 & \{0.444, 0.0494\}\end{array}$&\\ \hline
\reaction{p}{1/2}{d}{3/2}{d}{3/2}&$\begin{array}{ll}
 1 & \{0.543\} \\
 0 & \{0.84, 0.444\}\end{array}$&
\reaction{p}{3/2}{s}{1/2}{s}{1/2}&$\begin{array}{ll}
 3 & \{0\} \\
 2 & \{0, 0\} \\
 1 & \{ 0.543, 0, 0\} \\
 0 & \{0.84, 0.444, 0, 0\}\end{array}$&\\ \hline
\reaction{p}{3/2}{s}{1/2}{d}{3/2}&$\begin{array}{ll}
 3 & \{0\} \\
 2 & \{0.08, 0.00889\} \\
 1 & \{ 0.0622, 0.0491,0.00322\} \\
 0 & \{0.0494, 0.0178, 0, 0\}\end{array}$&
\reaction{p}{3/2}{s}{1/2}{d}{5/2}&$\begin{array}{ll}
 3 & \{0.267\} \\
 2 & \{0.48, 0.0533\} \\
 1 & \{0.64, 0.0533, 0.16\} \\
 0 & \{0.693, 0.267, 0, 0\}
\end{array}$&\\ \hline
\reaction{p}{3/2}{d}{3/2}{d}{3/2}&$\begin{array}{ll}
 3 & \{0.0128\} \\
 2 & \{0.00569, 0\} \\
 1 & \{0.00626, 0.00291, 0.00142\} \\
 0 & \{0.0178, 0.0149, 0.00217, 0.00182\}\end{array}$&
\reaction{p}{3/2}{d}{3/2}{d}{5/2}&$\begin{array}{ll}
 3 & \{0.0725\} \\
 2 & \{0.0672, 0.0587\} \\
 1 & \{0.0654, 0.0608, 0.0189\} \\
 0 & \{0.0675, 0.0657, 0.0328, 0.000416\}\end{array}$&\\ \hline
\reaction{p}{3/2}{d}{5/2}{d}{5/2}&$\begin{array}{ll}
3 & \{0.269\} \\
 2 & \{0.576, 0.269\} \\
 1 & \{ 0.831, 0.499, 0.264\} \\
 0 & \{0.935, 0.649, 0.428, 0.253\}
 \end{array}$&
\reaction{p}{3/2}{d}{}{d}{}&$
\begin{array}{ll}
 3 & \{0.354\} \\
 2 & \{0.635,0.342\} \\
 1 & \{0.857,0.561,0.33\} \\
 0 & \{0.948,0.699,0.501,0.321\}
\end{array}
$& \\ \hline\hline
\reaction{d}{3/2}{p}{1/2}{p}{1/2}&$\begin{array}{ll}
3 & \{0\} \\
 2 & \{0, 0\} \\
 1 & \{0.543, 0,  0\} \\
 0 & \{0.84, 0.444, 0, 0\}
  \end{array}$&
\reaction{d}{3/2}{p}{1/2}{p}{3/2}&$\begin{array}{ll}
 3 & \{0\} \\
 2 & \{0.08, 0.00889\} \\
 1 & \{0.0622, 0.0491, 0.00322\} \\
 0 & \{0.0494, 0.0178, 0, 0\}
  \end{array}$&\\ \hline
\reaction{d}{3/2}{p}{3/2}{p}{3/2}&$\begin{array}{ll}
3 & \{0.0128\} \\
 2 & \{0.00569, 0\} \\
 1 & \{0.00626, 0.00291, 0.00142\} \\
 0 & \{0.0178, 0.0149, 0.00217, 0.00182\}
  \end{array}$&
\reaction{d}{3/2}{p}{1/2}{f}{5/2}&$\begin{array}{ll}
 3 & \{0.267\} \\
 2 & \{0.48, 0.0533\} \\
 1 & \{0.64, 0.16, 0.0533\} \\
 0 & \{0.693, 0.267, 0, 0\}
  \end{array}$&\\ \hline
\reaction{d}{3/2}{p}{3/2}{f}{5/2}&$\begin{array}{ll}
 3 & \{0.0725\} \\
 2 & \{0.0672, 0.0587\} \\
 1 & \{0.0654, 0.0608, 0.0189\} \\
 0 & \{0.0675, 0.0657, 0.0328, 0.000416\}
  \end{array}$&
\reaction{d}{3/2}{f}{5/2}{f}{5/2}&$\begin{array}{ll}
 3 & \{0.269\} \\
 2 & \{0.576, 0.269\} \\
 1 & \{0.831, 0.499, 0.264\} \\
 0 & \{0.935, 0.649, 0.428, 0.253\}
  \end{array}$&\\ \hline
\reaction{d}{5/2}{p}{3/2}{p}{3/2}&$\begin{array}{ll}
5 & \{0\} \\
 4 & \{0, 0\} \\
 3 & \{0.269, 0, 0\} \\
 2 & \{0.576, 0.269, 0, 0\} \\
 1 & \{0.831, 0.499, 0.264, 0, 0\} \\
 0 & \{0.935, 0.649, 0.428, 0.253, 0, 0\}
  \end{array}$&
\reaction{d}{5/2}{p}{3/2}{f}{}&$\begin{array}{ll}
 5 & \{0.343\} \\
 4 & \{0.503, 0.137\} \\
 3 & \{0.642, 0.263, 0.0702\} \\
 2 & \{0.747, 0.373, 0.145, 0.0659\} \\
 1 & \{0.814, 0.442, 0.229, 0.0764, 0.0594\} \\
 0 & \{0.836, 0.466, 0.243, 0.152, 0.00304, 0.00239\}
  \end{array}$&\\ \hline 
\reaction{d}{5/2}{f}{}{f}{}&
\multicolumn{3}{l}{$\begin{array}{ll}
 5 & \{0.185\} \\
 4 & \{0.397, 0.193\} \\
 3 & \{0.619, 0.375, 0.201\} \\
  \end{array}\begin{array}{ll}2 & \{0.809, 0.561, 0.353, 0.206\} \\
 1 & \{0.938, 0.703, 0.505, 0.332, 0.208 \} \\
 0 & \{0.985, 0.763, 0.597, 0.458, 0.317, 0.207\}\end{array}$}
&\\ \hline
  \end{tabular}\end{table*}%

 \def\reaction#1#2#3#4#5#6{{#1_{#2}+#1_{#2}\rightarrow #3_{#4}+#5_{#6}}}

 We note that of the 23 different channels, 9 of them have at least 1 \forster\ zero state  with zero dipole-dipole interaction.  Another 7 have at least one state with an eigenvalue less than 0.05.  All of these states are extremely weakly coupled by dipole-dipole interactions, and therefore are of limited use for blockade experiments. 
  
 The remaining seven channels
 \begin{equation}
\begin{array}{c}
\reaction{s}{1/2}{p}{\;\;\;\;\;}{p}{\;\;\;\;\;} \\
\reaction{s}{1/2}{p}{3/2}{p}{3/2} \\
\reaction{p}{1/2}{d}{3/2}{d}{3/2}\\
\reaction{p}{3/2}{d}{5/2}{d}{5/2}\\
\reaction{p}{3/2}{d}{\;\;\;\;\;}{d}{\;\;\;\;\;}\\
\reaction{d}{3/2}{f}{5/2}{f}{5/2}\\
\reaction{d}{5/2}{f}{\;\;\;\;\;}{f}{\;\;\;\;\;}
\end{array}
\end{equation}
 have minimum eigenvalues of $0.18$ or greater.  All of them have the property that the state being coupled to by the dipole-dipole interaction has larger angular momentum than the initial state. This is consistent with the argument given above (Section~\ref{general}) and in Ref.~\cite{Walker05} that the preferred channels for blockade experiments have dominant intermediate channels where both atoms have  total angular momentum $j+1$.
  
The three channels with no fine structure specified in the final state deserve special mention. Each of them have two allowed intermediate angular momentum channels that contribute to the van der Waals interactions. When there is no near \forster\ resonance for one of those channels, to a good approximation the fine-structure in the final state can be neglected.  An example of this will be given in Section~\ref{sec:angle}.
    
The simplest case with zeros is $p_{1/2}+p_{1/2}\rightarrow s_{1/2}+s_{1/2}$.  In this case the $M=0$ portion of the $\mathcal D$ matrix is
\begin{equation}
\mathcal D= {8\over 81}\left(  
\begin{array}{cc}
      1 & 1 \\
    1 & 1\\
   \end{array}
   \right)
\end{equation}
which has eigenvalues of 0 and 16/81.  The zero eigenvector, $\ket{{1\over 2}{-1\over 2}}-\ket{{-1\over 2}{1\over 2}}$, has zero contribution to the van-der-Waals interaction from the $s+s$ states.

For the case of $p_{3/2}+p_{3/2}\rightarrow s_{1/2}+s_{1/2}$, the $M=\pm 2$ and $M=\pm 3$ states have zero dipole-dipole coupling since the $s+s$ states are limited to $M=0,\pm1$.  Only 4 of the 16 possible two-atom states have non-zero contributions to the van-der-Waals interaction from the $s+s$ channel.

\begin{figure}[t] 
   \includegraphics[width=3.0in]{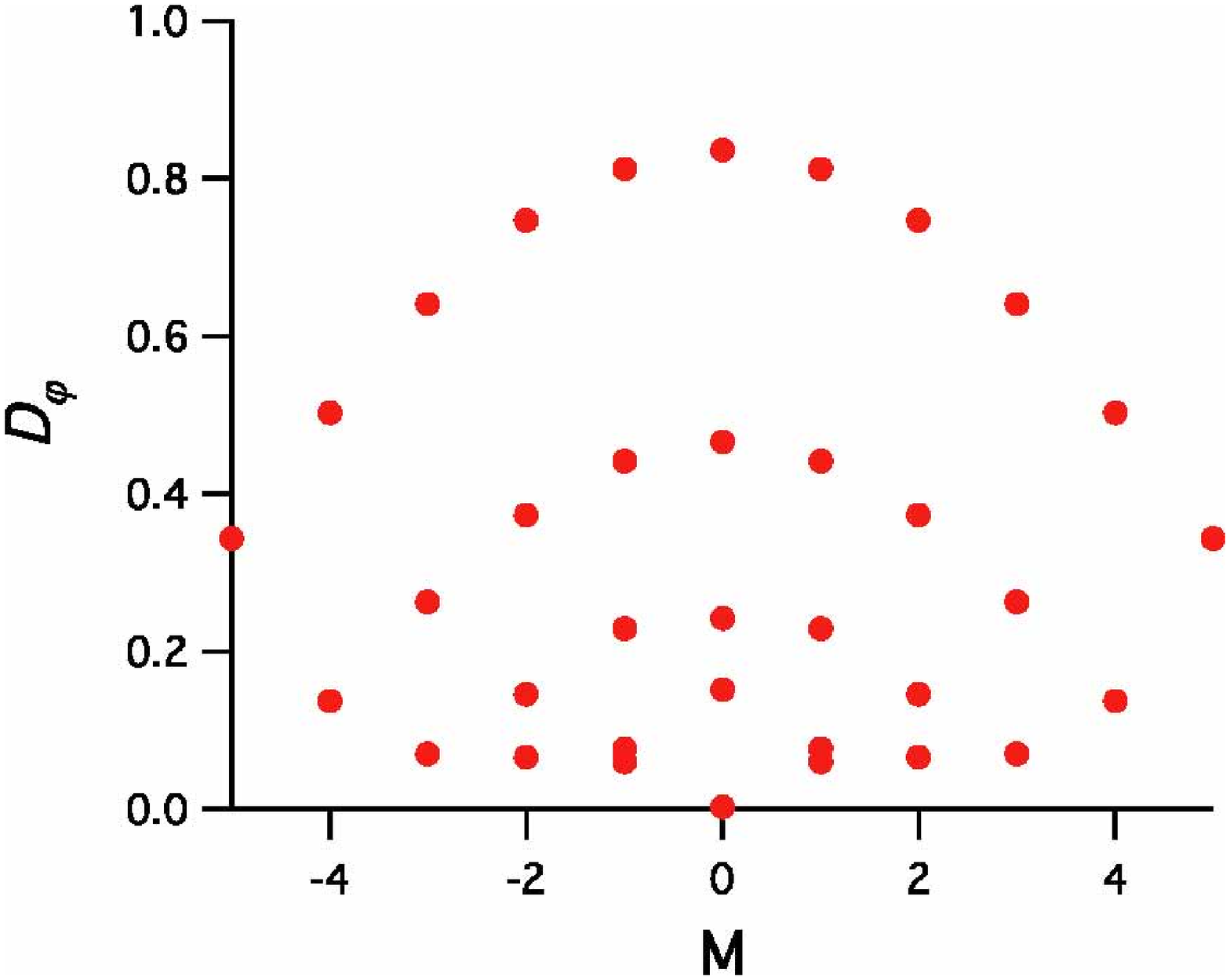} 
   \caption{(color online) Range of van-der-Waals coefficients for the channel $d_{5/2}+d_{5/2}\rightarrow p_{3/2}+ f_{5/2,7/2}$.  The $M=0$ point lying nearly on the $D_\varphi=0$ axis experiences little blockade. }
   \label{ddplot}
\end{figure}

An interesting case occurs in Rb due to the near resonance of $43d_{5/2}+43d_{5/2}\rightarrow 45p_{3/2}+41 f$.  The fine structure splitting of the f-states is small, so both the $f_{5/2}$ and $f_{7/2}$ states must be taken into account.  In the approximation of zero $f$ state fine structure splitting, we find that the eigenvalues of $\mathcal D$ range from  $0.836$ down to $0.0024$, a factor of 350.  The distribution of eigenvalues is shown in Figure~\ref{ddplot}.

Although the figure of merit for blockade is primarily determined by the energy shifts $\Delta_{\varphi kl}$, the angular distribution and polarization dependence can play an important role for specialized geometries.  These effects show up in  the overlap factor 
\begin{equation}
\kappa_{\varphi kl}={\rabi_{\gamma k}\rabi_{\gamma l}\over \rabi_{0}^{2}}\overlap{\varphi kl}{\gamma k\gamma l}
\end{equation}
where $\ket{{\gamma k\gamma l}}$ is the doubly-excited state that would be generated by the light in the absence of dipole-dipole interactions.
The wavefunctions $\ket{\varphi kl}$ are simplest when expressed in a coordinate system aligned with the interatomic separation, while $\ket{{\gamma k\gamma l}}$ is naturally represented in a fixed coordinate system.  It is therefore convenient to rotate the van der Waals eigenstates to the fixed frame, denoted by primes, to get
\begin{eqnarray}
\overlap{\varphi kl}{\gamma k\gamma l}&=&\sum_{\substack{m'_{k}m'_{l}\\m_{k}m_{l}}}\overlap{\varphi kl}{m_{k}m_{l}}d^{j}_{m_{k}m'_{k}}d^{j}_{m_{l}m'_{l}}\nonumber\\
&&\hspace*{0.5 in}\times \overlap{m'_{k}m'_{l}}{\gamma k\gamma l}
\end{eqnarray}
where the $d$'s are Wigner rotation matrices  evaluated at angle $\theta_{kl}$, the angle between the interatomic axis and the z-axis of the fixed coordinate system.

\subsection{Connection between van der Waals and \forster\ Regimes}

In the case that a single channel dominates, the eigenstates $\ket{\varphi}$ and eigenvalues $D_{\varphi}$ of $\mathcal M^{\dagger}\mathcal M$ can be used to analytically find the energies and eigenstates in the $R^{-3}$ \forster\ regime as well. The transition between van der Waals and F\"orster interactions occurs at a characteristic length scale of $R_c=(4 C_3^2/\delta^2)^{1/6}.$
A related analysis in a different context was given in Ref.~\cite{Zhang07}.

The $\mathcal M$ operator acting on a \forster\ eigenstate $\ket\varphi$ produces a unique vector $\ket{\chi_{\varphi}}$ that is a superposition of the Zeeman sublevels of the coupled state:
\begin{equation}
\mathcal M\ket\varphi=\sqrt{D_{\varphi}}\ket{\chi_{\varphi}}
\end{equation}
Operating on the left side with $\mathcal M\mathcal M^{\dagger}$ we get
\begin{eqnarray}
\mathcal M\mathcal M^{\dagger}\mathcal M\ket\varphi&=&\mathcal M D_{\varphi}\ket\varphi\\
 \mathcal M\mathcal M^{\dagger}\ket{\chi_{\varphi}}&=&D_{\varphi}\ket{\chi_{\varphi}}
\end{eqnarray}
so $\ket{\chi_{\varphi}}$ is an eigenvector of $\mathcal M\mathcal M^{\dagger}$ with eigenvalue $D_{\varphi}$.
It also follows that $\mathcal M^{\dagger}\ket{\chi_{\varphi}}=\sqrt{D_{\varphi}}\ket{\varphi}$, and $\langle\chi_{\varphi}\ket{\chi_{\varphi}}=1$.  Therefore the states $\ket\varphi$ and  $\ket{\chi_{\varphi}}$ form a closed two-level system under the influence of the dipole-dipole interaction.

It is now straightforward to find the eigenstates and eigenvalues in the \forster\ regime
as well.  The Hamiltonian matrix for the effective two-level system is
\begin{equation}
H_{\varphi}=
\left(
\begin{array}{cc}
\delta & {C_{3}\over R^{3}}\sqrt{D_{\varphi}}     \\
   {C_{3}\over R^{3}}\sqrt{D_{\varphi}}    &0           
\end{array}
\right)
\end{equation}
where  $\delta=E_{\varphi}-E_{\chi}$ is the \forster\ defect and $C_{6}=C_{3}^{2}/\delta$.  The eigenvalues are
\begin{equation}
V_{\pm}(R)={\delta\over 2}\pm\half1\sqrt{\delta^{2}+4C_{3}^{2}D_{\varphi}/R^{6}}
\end{equation}
and the eigenvectors are
\begin{equation}
\begin{array}{c}
\psi_{-}=\cos\theta\ket\varphi-\sin\theta\ket{\chi_{\varphi}}\\
\psi_{{+}}=\sin\theta\ket\varphi+\cos\theta\ket{\chi_{\varphi}}
\end{array}
\end{equation}
where $\tan2\theta=-2C_{3}\sqrt{D_{\varphi}}/(\delta R^{3})$.

These considerations show that the long-range potentials for $a+a\rightarrow b+b$ and the reversed $b+b\rightarrow a+a$ are anti-symmetric in energy about $\delta/2$.  If $j_{b}<j_{a}$, the channels with $|M|>2j_{b}$ have no dipole-dipole interaction.

In the limit of strong dipole-dipole coupling, we have 
\begin{equation}
V_{\pm}=\pm {C_{3}\over R^{3}}\sqrt{D_{\varphi}}
\end{equation}

\begin{figure}[t] 
   \centering
   \includegraphics[width=3.2in]{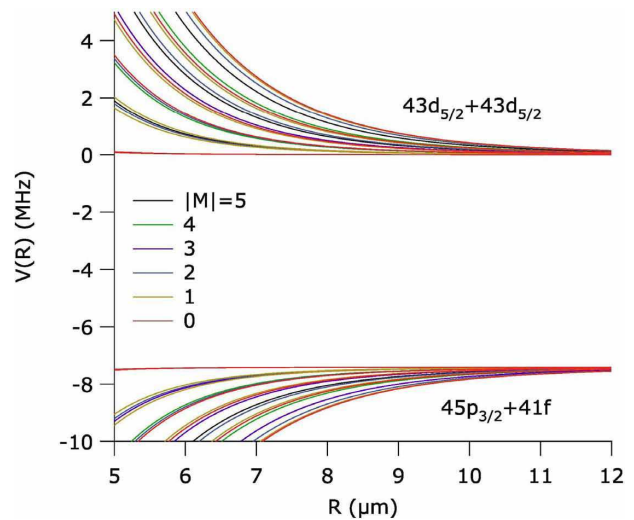} 
   \caption{(color online) The potential curves  for $\reaction{43d}{5/2}{45p}{3/2}{41}{f}$ in Rb, using $\delta=-7.4$ MHz and $C_3=1.98$ GHz $\mu$m$^3$ which give $R_c=8.1~\mu\rm m$.  At small $R$ additional channels contribute so these curves are not accurate there.}
   \label{vdWforster}
\end{figure}

The potential curves for $43d_{5/2}+43d_{5/2}\rightarrow45p_{3/2}+{41}{f}$, generated from Table~\ref{evals} and the analytical formula above,  are shown in Figure~\ref{vdWforster}.

\subsection{Evaluation of van der Waals Interactions for Rb and Cs Rydberg States}

\newcommand \be{\begin{equation}}
\newcommand \ee{\end{equation}}

We now proceed to quantitatively evaluate the van der Waals interactions of Rydberg states that can be reached by one or two photon   excitation from the ground state of neutral alkali atoms, with the restriction that we include only  cases where both atoms are initially excited to the same level.  For each choice of excited state we give numerical values for the energy defects and the interaction strength for $n=70$ as well as  for values of $n$ where resonances occur. Results are given for  the two heaviest alkali atoms Rb and Cs. Due to the large hyperfine splittings of the upperstates of the D1 and D2 lines in these atoms, they are the most promising candidates among the alkali metal atoms for quantum logic experiments which rely on a well resolved excited state hyperfine structure for qubit initialization and readout. 

Before discussing  the cases individually  we recall that the 
   long range interaction strength of a particular channel scales proportional to $C_{6}=\left(\rmat{\gamma_{i}}{\gamma_{s}}\rmat{\gamma_i}{\gamma_{t}}\right)^2/|\delta|$
where we have introduced a shorthand notation $\gamma=\{nlj\}$ for the quantum numbers  specifying the initial laser excited $(\gamma_i)$ and F\"orster coupled  $(\gamma_{s},\gamma_{t})$ states. We have calculated these matrix elements in several ways: using model potentials\cite{Marinescu94},  with  quantum defect wavefunctions\cite{Burgess60}, and using a semiclassical analytical formula\cite{Kaulakys95}.  Results from the quantum defect wavefunctions and semiclassical calculations typically agree to better than 1\%, while the model wavefunction calculations differ from these by up to 10\%.  The numerical values in what follows  were obtained  from the quantum defect wavefunction calculations. {The radial integrals depend explicitly on $n,l$ but also have an implicit dependence on $j$  due to the dependence of the quantum defects on the fine structure level. This leads to a $j$ dependence of as much as 10\% in some cases. We report numerical values for the radial integrals corresponding to the particular fine structure channel considered.  }

\begin{figure}
\centering
\includegraphics[width=3.3 in]{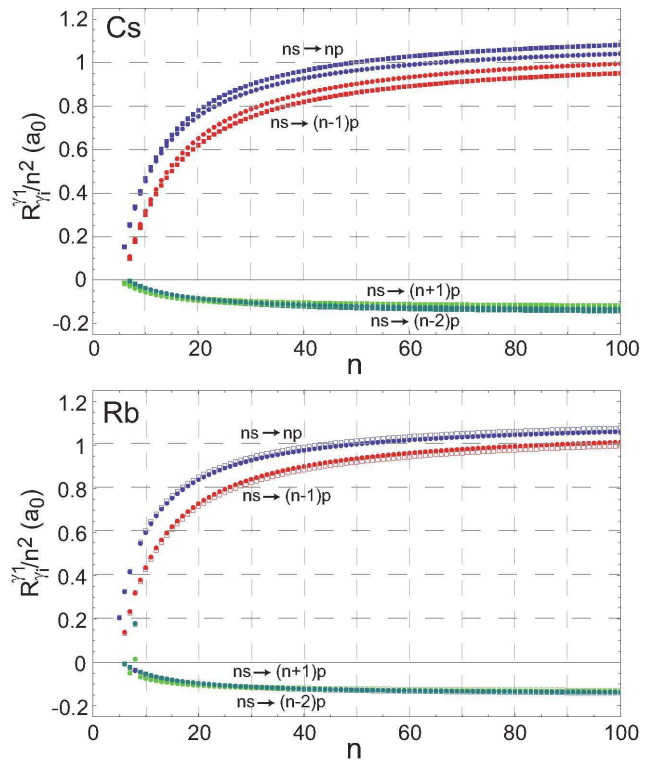}
  \caption{(color online) Radial matrix elements divided by $n^2$ for transitions
 $ns_{1/2} \rightarrow  n_sp_{3/2}$ (filled circles) and 
$ns_{1/2}\rightarrow n_sp_{1/2}$ (filled boxes) in Cs and Rb.  
 }
\label{fig.spradial}
\end{figure}

\begin{figure}[!t]
\centering
\includegraphics[width=3.3 in]{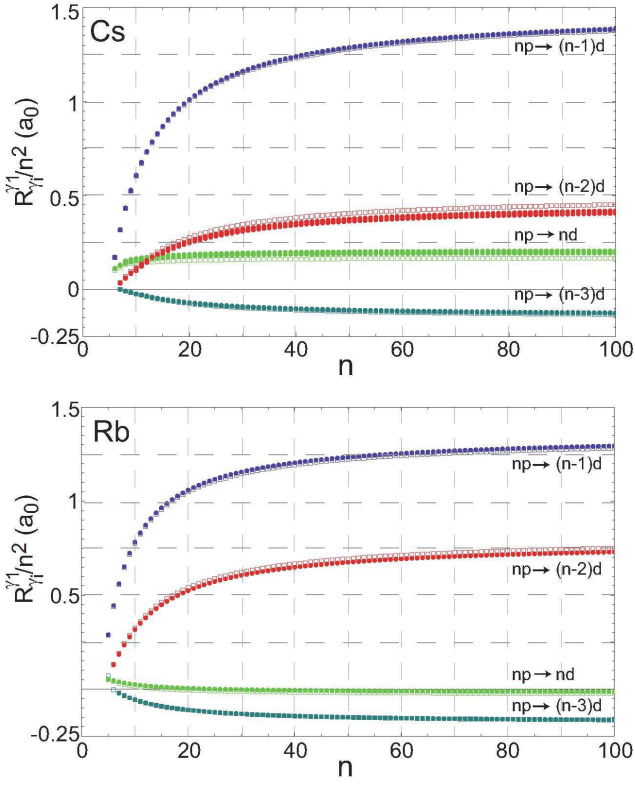}
  \caption{(color online) Radial matrix elements divided by $n^2$ for transitions
 $np_{1/2} \rightarrow  n_sd_{3/2}$ (open boxes) and 
$np_{3/2}\rightarrow n_sd_{3/2}$ (filled circles) in Cs and Rb.
In Cs the additional fine structure transitions $np_{3/2}\rightarrow n_sd_{5/2}$
 are also shown with filled circles and have  slightly smaller matrix elements than the transitions to $d_{3/2}$ for $n_s=n$ and $n_s=n-1$, and slightly larger matrix elements than the transitions to $d_{3/2}$ for $n_s=n-2$ and $n_s=n-3$. In Rb the differences between $d_{3/2}$ and $d_{5/2}$ are less than 1\% and are not shown. }
\label{fig.pdradial}
\end{figure}

\begin{figure}[!t]
\centering
\includegraphics[width=3.3 in]{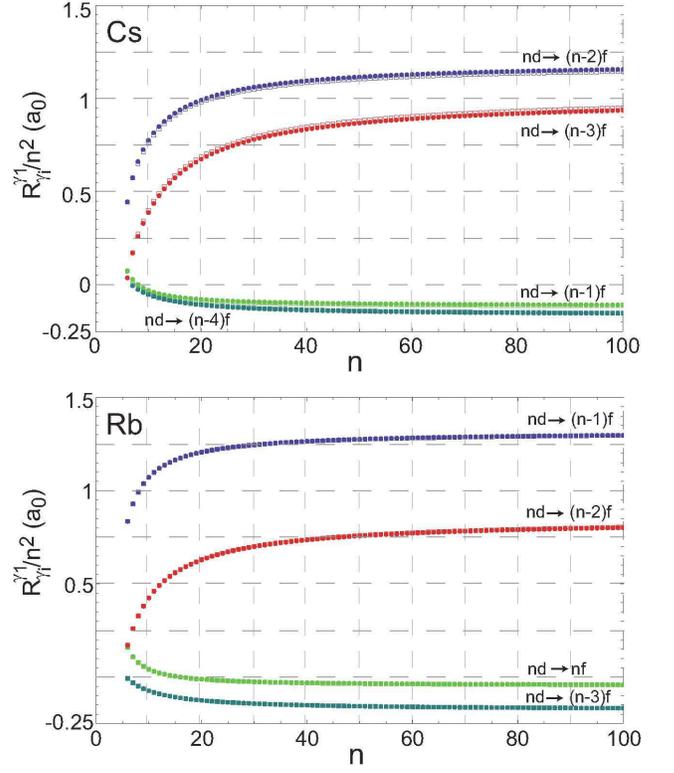}
  \caption{(color online) Radial matrix elements divided by $n^2$ for transitions
 $nd_{3/2} \rightarrow  n_sf_{5/2}$ (open boxes) and 
$nd_{5/2}\rightarrow n_sf_{5/2}$ (filled circles) in Cs and Rb.
The additional fine structure transitions $nd_{5/2}\rightarrow n_sf_{7/2}$ differ by less than 0.1\% from the $f_{5/2}$ case. 
 }
\label{fig.dfradial}
\end{figure}

In Figs. \ref{fig.spradial}, \ref{fig.pdradial}, and \ref{fig.dfradial}
we show the scaled radial matrix elements for all the dipole allowed transitions between $s$, $p$, $d$ and $f$ Rydberg states in Cs and Rb.
The radial matrix elements between high lying Rydberg levels are strongly
peaked for states of similar energy so in practice only small positive or negative values of $n_s-n_{i}$ occur.   
The calculations reveal that in all cases  the matrix elements are large for at most three 
values of $n_s-n_{i}$. This is important since it limits the number of F\"orster channels which must be taken into account for an accurate calculation of the interaction strength.
 The matrix elements are close to their asymptotic $n^2 a_0$ scaling   
for $n>50$ and fine structure dependent effects are generally small except for   the 
$s\rightarrow p$ and $p\rightarrow d$ transitions in Cs, which has a larger fine structure splitting than Rb.

The energy defects $\delta$ were calculated using  recently measured values for the Rb quantum defects \cite{Li03,Han06}, and older data for Cs\cite{Lorenzen84}.  Since the radial integrals scale as $n^2$ and the energy defects scale as $n^{-3}$ the interaction strength usually scales as $C_6\sim n^{11}.$ As we will see below  the asymptotic $n^{11}$ scaling is often broken for specific values of $n<100$ where the 
quantum defects conspire to give near resonant F\"orster interactions. 
These special values of $n$ may be particularly useful for engineering strong interactions without needing to access very high lying states. 

\subsubsection{$ns_{1/2} + ns_{1/2}\leftrightarrow n_sp_j+ n_tp_j$}

The first particular case is the excitation of $ns_{1/2}$ states which can F\"orster couple to pairs of $n_sp_{j},n_tp_j$ states with $j=1/2,3/2.$ There are three possible fine structure channels  giving F\"orster defects
\bse 
\bea
\hspace*{-0.2 in}\delta_1(n_s,n_t)&=&E(n_sp_{3/2})+E(n_tp_{3/2})-2E(ns_{1/2})
\\
\hspace*{-0.2 in}\delta_2(n_s,n_t)&=&E(n_sp_{3/2})+E(n_tp_{1/2})-2E(ns_{1/2})
\\
\hspace*{-0.2 in}\delta_3(n_s,n_t)&=&E(n_sp_{1/2})+E(n_tp_{1/2})-2E(ns_{1/2})
\eea
\label{stopchannels}
\ese

\begin{figure}[!t]
\centering
\includegraphics[width=3.3 in]{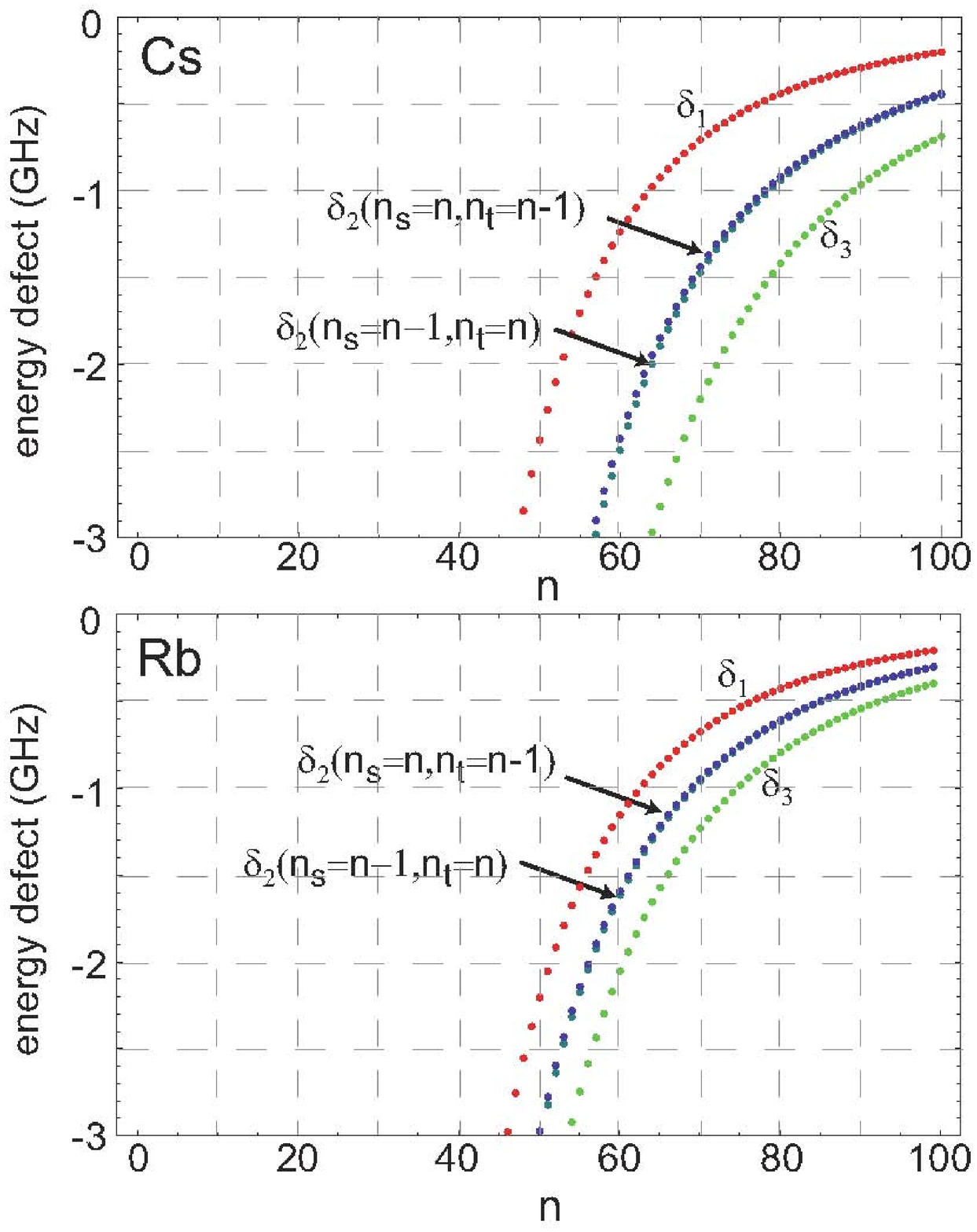}
  \caption{(color online) F\"orster energy defects for $ns_{1/2} \leftrightarrow  n_sp_j+n_tp_j$ coupling in Cs and Rb.
 }
\label{fig.spdefect}
\end{figure}

The situation for Cs and Rb is shown in Fig.  \ref{fig.spdefect}.
The behavior is similar for both  with the energy defect decreasing like $1/n^3$ although the larger fine structure splitting in Cs separates the $p_{1/2}$ and $p_{3/2}$ channels as compared to  Rb. At $n=70$ in Cs the strongest channels are $\delta_1$(70,69), $\delta_2$(70,69), $\delta_2$(69,70), and $ \delta_3$(70,69), giving $C_{6}=716, 315, 381, 227 ~\rm GHz~\mu m^6$.  The next contribution is that from   
$\delta_1(71,68)/2\pi=-2.8~\rm GHz$ which has much smaller radial matrix elements giving $C_{6}=0.05~\rm GHz~\mu m^6$. 
For Rb we find that $\delta_1(70,69), \delta_2(70,69), \delta_2(69,70),$ and $ \delta_3(70,69)$ give $C_{6}=799,  543, 589, 437 ~\rm GHz~\mu m^6$. The next contribution is that from   
$\delta_1(71,68)/2\pi=-2.69~\rm GHz$ which has much smaller radial matrix elements giving $C_{6}=0.06~\rm GHz~\mu m^6$.  As we will see in Section \ref{sec:angle}  the variation of the energy defects between channels results in an almost isotropic interaction for Rb, and some slight angular variation for Cs.

\subsubsection{$np_j + np_j\leftrightarrow n_ss_{1/2}+ n_ts_{1/2}$}

\begin{figure}[!tb]
\centering
\includegraphics[width=3.3 in]{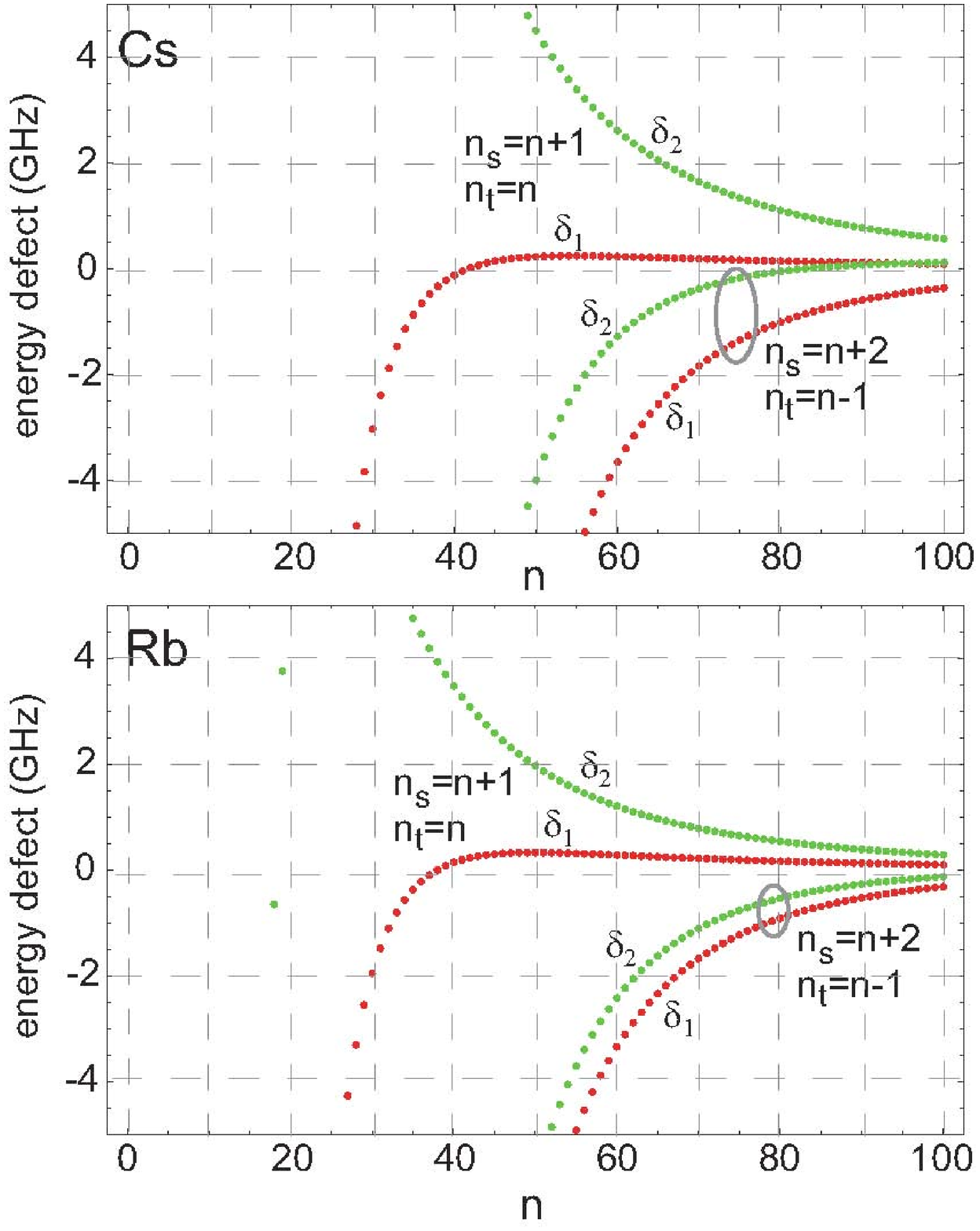}
  \caption{(color online) F\"orster energy defects for $np_j \leftrightarrow  n_ss_{1/2}+n_ts_{1/2}$ coupling in Cs and Rb.
 }
\label{fig.psdefect}
\end{figure}

The next case is coupling of $p_j$ states with $j=1/2,3/2$ to $s_{1/2}$ states. 
There are two  fine structure channels with energy defects
\bse\bea
\delta_1&=&E(n_ss_{1/2})+E(n_ts_{1/2})-2E(np_{3/2}) \nonumber
\\
\delta_2&=&E(n_ss_{1/2})+E(n_ts_{1/2})-2E(np_{1/2})\nonumber
\eea\ese
which are shown in Fig. \ref{fig.psdefect}.

For Cs with $n_s=n+1, n_t=n$  channel 1 has a  resonance at $n=42$ where $\delta_1/2\pi=15.7~\rm MHz$ and
the corresponding interaction strength is $C_6= -432~\rm GHz~\mu m^6$. The high $n$ interaction strength in this channel is $C_6= -2920~\rm GHz~\mu m^6$ at $n=70.$ The $\delta_2$ channel for $n_s=n+1, n_t=n$ is substantially weaker giving  $C_6= -324~\rm GHz~\mu m^6$ at $n=70.$
The $\delta_2$ channel   also has a high $n$ resonance 
for $n_s=n+2, n_t=n-1$ with $\delta_2/2\pi=-5.77~\rm MHz$ at $n=83$. However the matrix elements are small so we get a relatively weak  interaction of $C_6= 104~\rm GHz~\mu m^6$. 

For Rb with $n_s=n+1, n_t=n$  channel 1 has a   resonance at $n=38$ where $\delta_1/2\pi=-4.1~\rm MHz$
the corresponding interaction strength is $C_6= 843~\rm GHz~\mu m^6$. At  $n=70,$ $n_s=n+1, n_t=n,$  the $\delta_1,\delta_2$ channels give $C_6= -2820$ and $ -767~\rm GHz~\mu m^6.$ 

We see that both species have resonances which provide a strong interaction at relatively low $n$ in addition to very strong interactions at high $n$. Unfortunately the strongest interaction occurs in the $\delta_1$ channel which has angular zeroes (see Table \ref{evals}) so it is only useful for special geometries where the zeroes can be avoided.

\subsubsection{$np_j + np_j\leftrightarrow n_ss_{1/2}+ n_td_{j_2}$}

\begin{figure}[!tb]
\centering
\includegraphics[width=3.3 in]{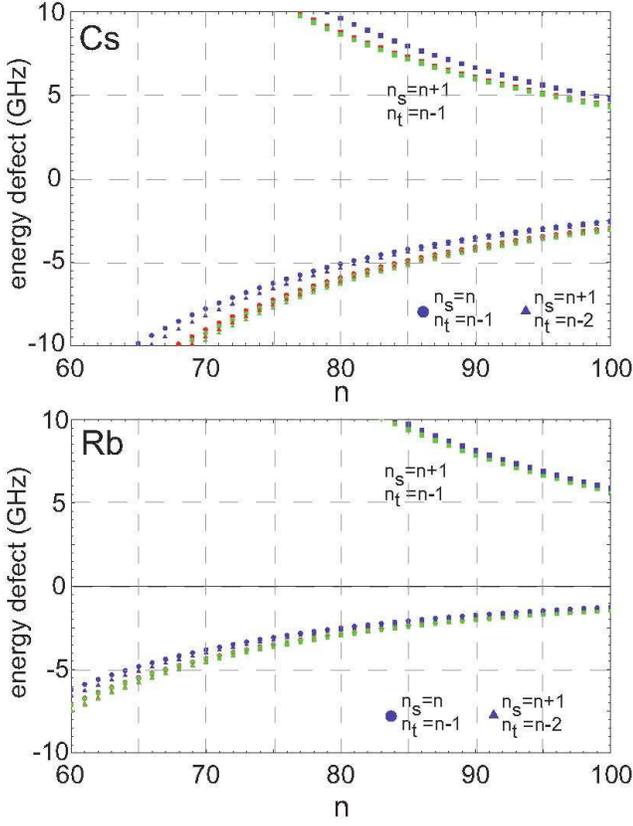}
  \caption{(color online) F\"orster energy defects for $np_j \leftrightarrow  n_ss_{1/2}+n_td_{j'} $ coupling in Cs and Rb.
 For the cases with negative energy defect the defect is smallest in magnitude for the $\delta_3$ channel, and for the the case with positive energy defect the $\delta_2$ channel has the defect with  smallest magnitude.
The $\delta_1$ and $\delta_2$ channels are very close together due to the small fine-structure splitting of the $d$ states.}
\label{fig.psddefect}
\end{figure}

The next case is coupling of $p_j$ states with $j=1/2,3/2$ to $s_{1/2}$ and $d_{j_2}$ states
with $j_2=3/2,5/2$. 
There are three  fine structure channels with energy defects
\bse\bea
\delta_1&=&E(n_ss_{1/2})+E(n_td_{5/2})-2E(np_{3/2}) \nonumber
\\
\delta_2&=&E(n_ss_{1/2})+E(n_td_{3/2})-2E(np_{3/2}) \nonumber
\\
\delta_3&=&E(n_ss_{1/2})+E(n_td_{3/2})-2E(np_{1/2}) \nonumber
\eea\ese
which are shown in Fig. \ref{fig.psddefect}. The radial matrix elements are large for $n_s=n, n+1$ and $n_t=n-2,n-1$, so we focus on these cases.

For Cs the three cases  $n_s=n$, $n_t=n-1$, $n_s=n+1$, $n_t=n-2$ and $n_s=n+1$, $n_t=n-1$ lie within a factor of 10 in strength for all fine structure channels. At $n=70$ we find for $C_6(\delta,n_s,n_t)$:
$$
\begin{array}{ll}
C_6(\delta_1,70,69)=111;
&C_6(\delta_2,70,69)=109\\
C_6(\delta_3,70,69)= 137;
&C_6(\delta_1,71,68)= 8.95 \\
C_6(\delta_2,71,68)= 8.2;
&C_6(\delta_3,71,68)= 11.0\\
C_6(\delta_1,71,69)= -71.5;
&C_6(\delta_2,71,69)= -72.9\\
C_6(\delta_3,71,69)= -59.0\\
\end{array}
$$ all in units of $~\rm GHz~\mu m^6.$

In Rb the behavior is similar except the $n_s=n+1$, $n_t=n-1$ case has a larger energy de\-fect than the others. 
At $n=70$ we find for $C_6(\delta,n_s,n_t):$
$$\begin{array}{ll}
C_6(\delta_1,70,69)= 218;
&C_6(\delta_2,70,69)= 217\\
C_6(\delta_3,70,69)= 253&
C_6(\delta_1,71,68)= 61.9\\
C_6(\delta_2,71,68)= 61.0&
C_6(\delta_3,71,68)= 71.0\\
C_6(\delta_1,71,69)= -52.4&
C_6(\delta_2,71,69)= -52.6\\  
C_6(\delta_3,71,69)= -48.2&\\
\end{array}$$
all in units of $~\rm GHz~\mu m^6.$
As can be seen in Table \ref{evals} the $\delta_1, \delta_2$ channels suffer from zero eigenvalues, but the $\delta_3$ channel does not and is therefore a good candidate for blockade experiments.

\subsubsection{$np_j + np_j\leftrightarrow n_sd_{j_1}+ n_td_{j_2}$}

\begin{figure}[tb]
\centering
\includegraphics[width=3.3 in]{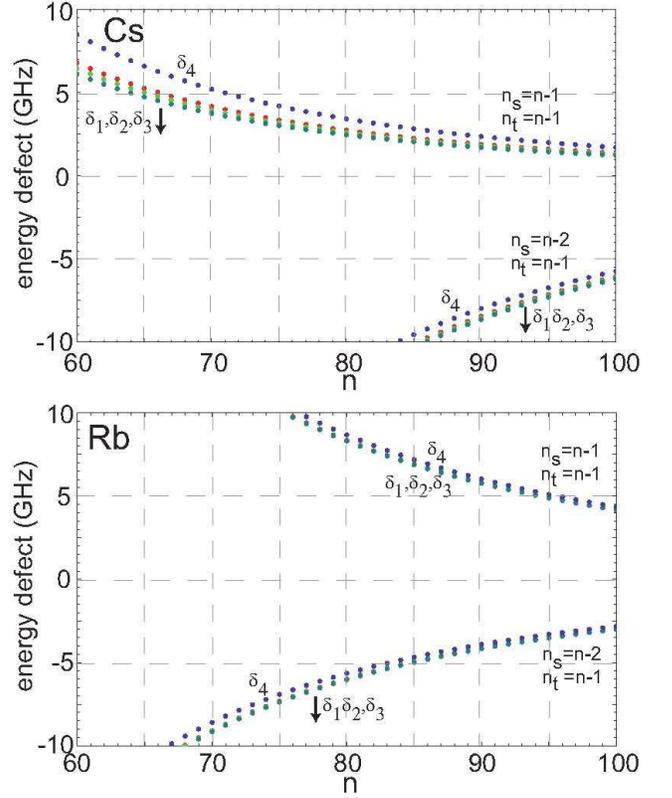}
  \caption{(color online) Energy defects for $np_j + np_j\leftrightarrow n_sd_{j_1}+ n_td_{j_2}$ coupling in Cs and Rb.
 }
\label{fig.pddefect}
\end{figure}

The next case is coupling of $p_j$ states with $j=1/2,3/2$ to $d_{j'}$ states
with $j'=3/2,5/2$. 
There are four fine-structure channels with energy defects
\bea
\delta_1&=&E(n_sd_{5/2})+E(n_td_{5/2})-2E(np_{3/2})\nonumber
\\
\delta_2&=&E(n_sd_{5/2})+E(n_td_{3/2})-2E(np_{3/2})\nonumber
\\
\delta_3&=&E(n_sd_{3/2})+E(n_td_{3/2})-2E(np_{3/2})\nonumber
\\
\delta_4&=&E(n_sd_{3/2})+E(n_td_{3/2})-2E(np_{1/2}).\nonumber
\eea

In Cs the strongest interactions occur for  $n_s=n-1$, $n_t=n-1$, and $n_s=n-2$, $n_t=n-1$ and the corresponding energy defects are shown in 
Fig. \ref{fig.pddefect}. 
At $n=70$ we find for $C_6(\delta,n_s,n_t):$ 
$$\begin{array}{ll}
C_6(\delta_1,69,69)= -428.;&
C_6(\delta_2,69,69)= -451.\\
C_6(\delta_3,69,69)= -478.;&
C_6(\delta_4,69,69)= -334.\\
C_6(\delta_1,68,69)= 8.42;&
C_6(\delta_2,68,69)= 8.36\\
C_6(\delta_3,68,69)= 7.75;& 
C_6(\delta_4,68,69)= 10.3\\ 
\end{array}$$
all in units of $~\rm GHz~\mu m^6.$ There is also a resonance at $n=68$ for $n_s=n+1, n_t=n-3$ for which 
the energy defect is remarkably small, $\delta_1/2\pi= 0.57~\rm MHz.$
  This value should be considered 
approximate since the quantum defects are not known accurately enough to predict the resonance to better than a  few MHz. 
The smallest of the other resonances at the same value of $n$ is 
$\delta_2/2\pi= -250~\rm MHz$ so to a good approximation we can consider just the first channel. The radial matrix elements are  $\rmat{69d}{68p}=-352~ a_0$ and $\rmat{65d}{68p}=-553~ a_0$ giving   
$C_6=-63.5 ~\rm GHz~\mu m^6.$ It should be noted that because the defect is so small the crossover to van der Waals behavior occurs at $R_c = 8.7~\mu\rm m$ despite the relatively weak strength of the interaction. 
 
In Rb   the strongest interactions also occur for  $n_s=n-1, n_t=n-1$, and $n_s=n-2, n_t=n-1$ . 
At $n=70$ we find for $C_6(\delta,n_s,n_t):$
$$\begin{array}{ll}
C_6(\delta_1,69,69)= -112. ;&
C_6(\delta_2,69,69)= -113. \\
C_6(\delta_3,69,69)= -113.;&
C_6(\delta_4,69,69)= -104.\\
C_6(\delta_1,68,69)= 48.6;&
C_6(\delta_2,68,69)= 48.6\\
C_6(\delta_3,68,69)= 48.0;& 
C_6(\delta_4,68,69)= 53.1 \\
\end{array}$$
all in units of $~\rm GHz~\mu m^6.$

\subsubsection{$nd_{3/2,5/2} + nd_{3/2,5/2}\leftrightarrow n_sp_{1/2,3/2}+ n_tp_{1/2,3/2}$}

\begin{figure}[tb]
\centering
\includegraphics[width=3.3 in]{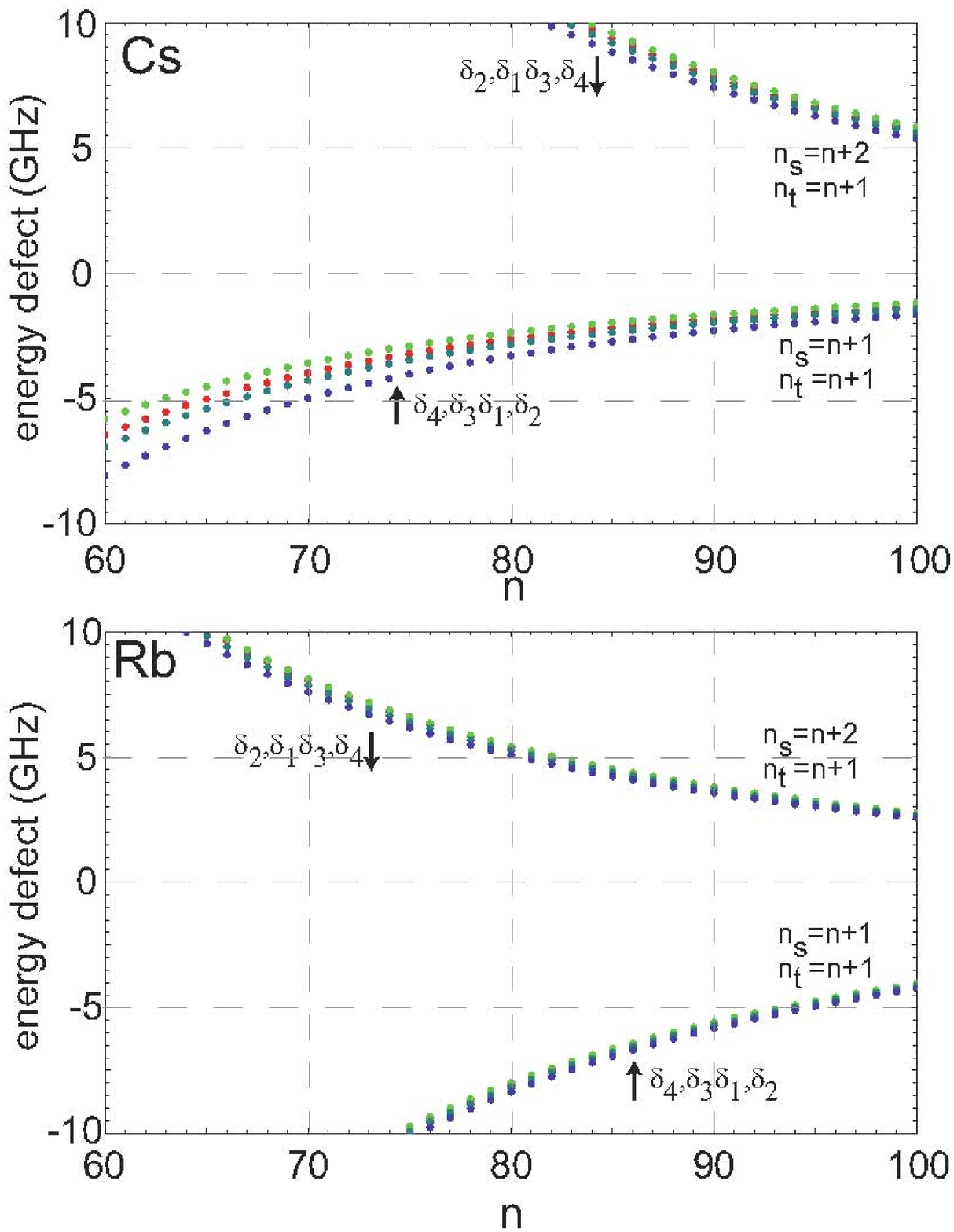}
  \caption{(color online) Energy defects for $nd_{3/2,5/2} + nd_{3/2,5/2}\leftrightarrow n_sp_{1/2,3/2}+ n_tp_{1/2,3/2}$ coupling in Cs and Rb.
 }
\label{fig.dpdefect}
\end{figure}

The final groups of cases to consider are the $d_{3/2},d_{5/2}$ states. Coupling of $d_j\leftrightarrow p_{j'}$ 
occurs for four possible channels
\bse\bea
\delta_1&=&E(n_sp_{3/2})+E(n_tp_{3/2})-2E(nd_{5/2})\nonumber
\\
\delta_2&=&E(n_sp_{3/2})+E(n_tp_{3/2})-2E(nd_{3/2})\nonumber
\\
\delta_3&=&E(n_sp_{3/2})+E(n_tp_{1/2})-2E(nd_{3/2})\nonumber
\\
\delta_4&=&E(n_sp_{1/2})+E(n_tp_{1/2})-2E(nd_{3/2})\nonumber
\eea\ese
which are shown in Fig. \ref{fig.dpdefect}.

In Cs the strongest channels are $n_s=n+1,n_t=n+1$ and $n_s=n+2,n_t=n+1$. 
At $n=70$ we find for $C_6(\delta,n_s,n_t):$ $$\begin{array}{ll}
C_6(\delta_1,71,71)= 504. ;&
C_6(\delta_2,71,71)= 563. \\
C_6(\delta_3,71,71)= 464.;&
C_6(\delta_4,71,71)= 394.\\
C_6(\delta_1,71,72)= -11.1;&
C_6(\delta_2,71,72)= -10.2\\
C_6(\delta_3,71,72)= -13.2;&
C_6(\delta_4,71,72)= -13.5 \end{array}$$
all in units of $~\rm GHz~\mu m^6.$
The same channels dominate in Rb where we find $$\begin{array}{ll}
C_6(\delta_1,71,71)= 132. ;&
C_6(\delta_2,71,71)= 133. \\
C_6(\delta_3,71,71)= 127.;&
C_6(\delta_4,71,71)= 122.\\
C_6(\delta_1,71,72)= -65.4;&
C_6(\delta_2,71,72)= -64.8\\
C_6(\delta_3,71,72)= -70.5;&
C_6(\delta_4,71,72)= -71.7
\end{array}$$
 in units of $~\rm GHz~\mu m^6.$
In this case there is a large difference between the species with the total interaction strength summed over the four channels about eight times larger in Cs than Rb. However, all of these channels have zero eigenvalues and are therefore not immediately useful for blockade.

\subsubsection{$nd_{3/2,5/2} + nd_{3/2,5/2}\leftrightarrow n_sp_{1/2,3/2}+ n_tf_{5/2,7/2}$}

\begin{figure}[!tb]
\centering
\includegraphics[width=3.3 in]{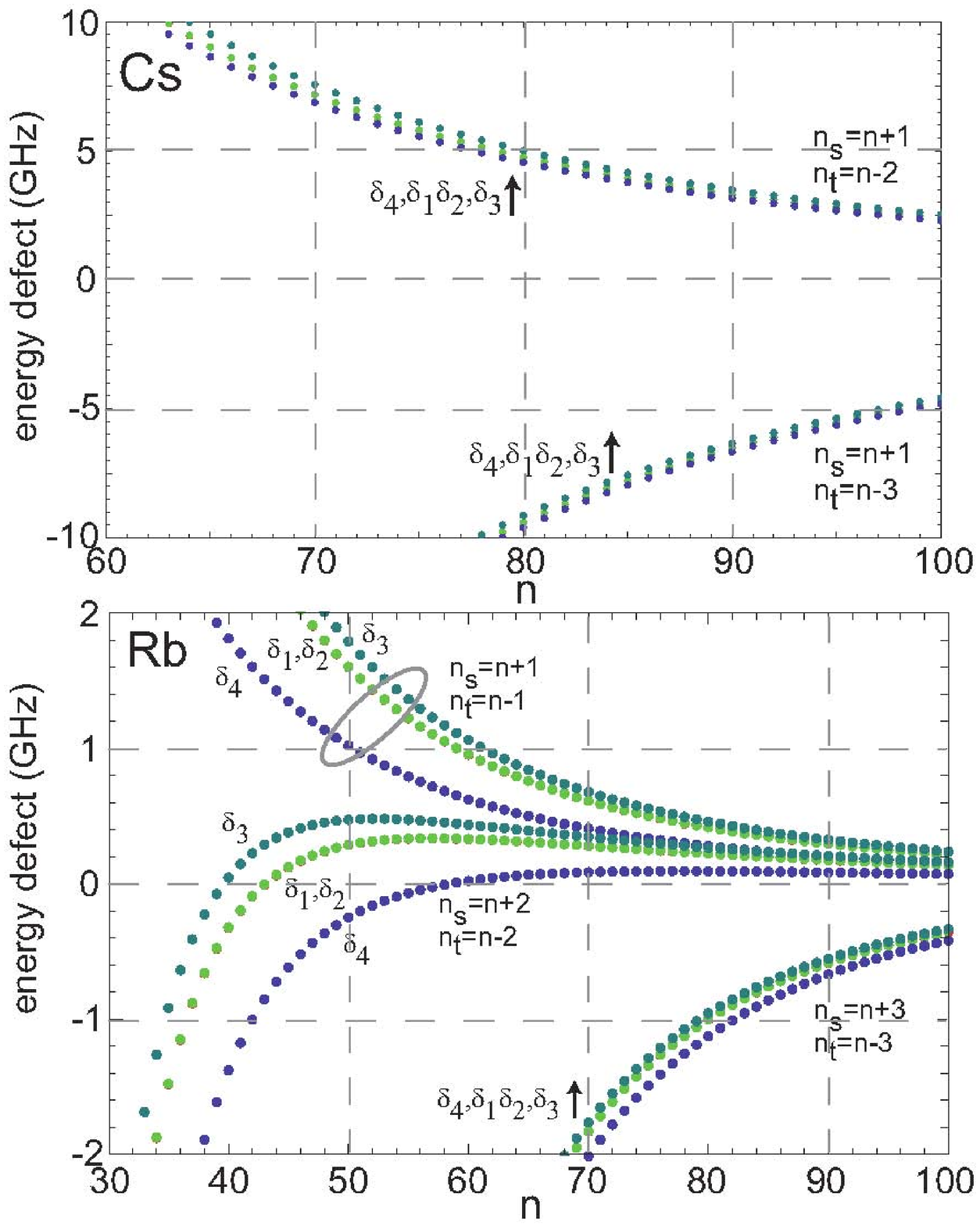}
  \caption{(color online) Energy defects for $nd_{3/2,5/2} + nd_{3/2,5/2}\leftrightarrow n_sp_{1/2,3/2}+ n_tf_{5/2,7/2}$ coupling in Cs and Rb.
 }
\label{fig.dpfdefect}
\end{figure}

The next case is coupling of $nd_j+nd_j\leftrightarrow n_sp_{j_1}+n_tf_{j_2}$. 
There are four possible channels
\bse\bea
\delta_1&=&E(n_sp_{3/2})+E(n_tf_{7/2})-2E(nd_{5/2})
\\
\delta_2&=&E(n_sp_{3/2})+E(n_tf_{5/2})-2E(nd_{5/2})
\\
\delta_3&=&E(n_sp_{3/2})+E(n_tf_{5/2})-2E(nd_{3/2})
\\
\delta_4&=&E(n_sp_{1/2})+E(n_tf_{5/2})-2E(nd_{3/2})
\eea
\label{pdfchannels}
\ese
which are shown in Fig. \ref{fig.dpfdefect}.

In Cs the channels  $n_s=n+1,n_t=n-3$ and $n_s=n+1,n_t=n-2$ have comparable interaction strengths. 
At $n=70$ we find for $C_6(\delta,n_s,n_t):$
$$\begin{array}{ll}
C_6(\delta_1,71,67)= 60.5;&
C_6(\delta_2,71,67)= 60.5\\
C_6(\delta_3,71,67)= 64.3;&
C_6(\delta_4,71,67)= 60.3\\
C_6(\delta_1,71,68)= -190.;&
C_6(\delta_2,71,68)= -190.\\
C_6(\delta_3,71,68)= -177.;&
C_6(\delta_4,71,68)= -192.
\end{array}$$ 
in units of $~\rm GHz~\mu m^6.$ There are other cases such as $n_s=n+4, n_t=n-5$ which have resonances, e.g. $\delta_4/2\pi = -39~\rm MHz$ at $n=65.$ However the radial matrix elements are too small to be useful.

In Rb the strongest channel at large $n$ is $n_s=n+1, n_t=n-1$, withÊ $n_s=n+2, n_t=n-2$ contributing about 30\% as large a $C_6$ andÊÊ $n_s=n+3, n_t=n-3$ being substantially weaker.Ê 
At $n=70$ we find for $C_6(\delta,n_s,n_t):$
$$\begin{array}{ll}
C_6(\delta_1,71,69)= -2530 ;&
C_6(\delta_2,71,69)= -2530 \\
C_6(\delta_3,71,69)= -2280 ;& 
C_6(\delta_4,71,69)= -3740  \\
C_6(\delta_1,72,68)= -677 ;&
C_6(\delta_2,72,68)= -676\\
C_6(\delta_3,72,68)= -547 ;&
C_6(\delta_4,72,68)= -2330.
\end{array}$$
in units of $~\rm GHz~\mu m^6.$

The $n_s=n+2, n_t=n-2$ channel is particularly interesting as it has a near resonance at $n=43$ where 
 $\delta_1/2\pi=-8.3~\rm MHz$ and  $\delta_2/2\pi=-6.0~\rm MHz$ as well as $n=58$ where 
$\delta_4/2\pi=-6.9~\rm MHz$ and $n=59$ where $\delta_4/2\pi=8.6~\rm MHz.$
The corresponding interaction strengths are 
$C_6(\delta_1,45,41)= 391$,
$C_6(\delta_2,45,41)= 539$, 
$C_6(\delta_4,60,56)= 6090$, and 
$C_6(\delta_4,61,57)= -5680$, 
in $~\rm GHz~\mu m^6.$  The $n=58$ resonance is the strongest we have found for any states with 
$n<70$ and has a crossover length of 
$R_c=11.8~\mu\rm m.$  Unfortunately the interaction has angular zeroes (see Table \ref{evals}) and is only useful for special geometries as discussed in Sec. \ref{sec:angle}.

\subsubsection{$nd_{3/2,5/2} + nd_{3/2,5/2}\leftrightarrow n_sf_{5/2,7/2}+ n_tf_{5/2,7/2}$}

The final case is coupling of $nd_{j}\leftrightarrow n_sf_{j_1}+n_tf_{j_2}.$ 
There are four possible channels  to consider
\bse\bea
\delta_1&=&E(n_sf_{7/2})+E(n_tf_{7/2})-2E(nd_{5/2})
\\
\delta_2&=&E(n_sf_{7/2})+E(n_tf_{5/2})-2E(nd_{5/2})
\\
\delta_3&=&E(n_sf_{5/2})+E(n_tf_{5/2})-2E(nd_{5/2})
\\
\delta_4&=&E(n_sf_{5/2})+E(n_tf_{5/2})-2E(nd_{3/2}).
\eea\ese
The energy defects for these channels are all very similar due to the smallness of the fine-structure splitting of the $d$ and $f$ states.

\begin{figure}[!tb]
\centering
\includegraphics[width=3.3 in]{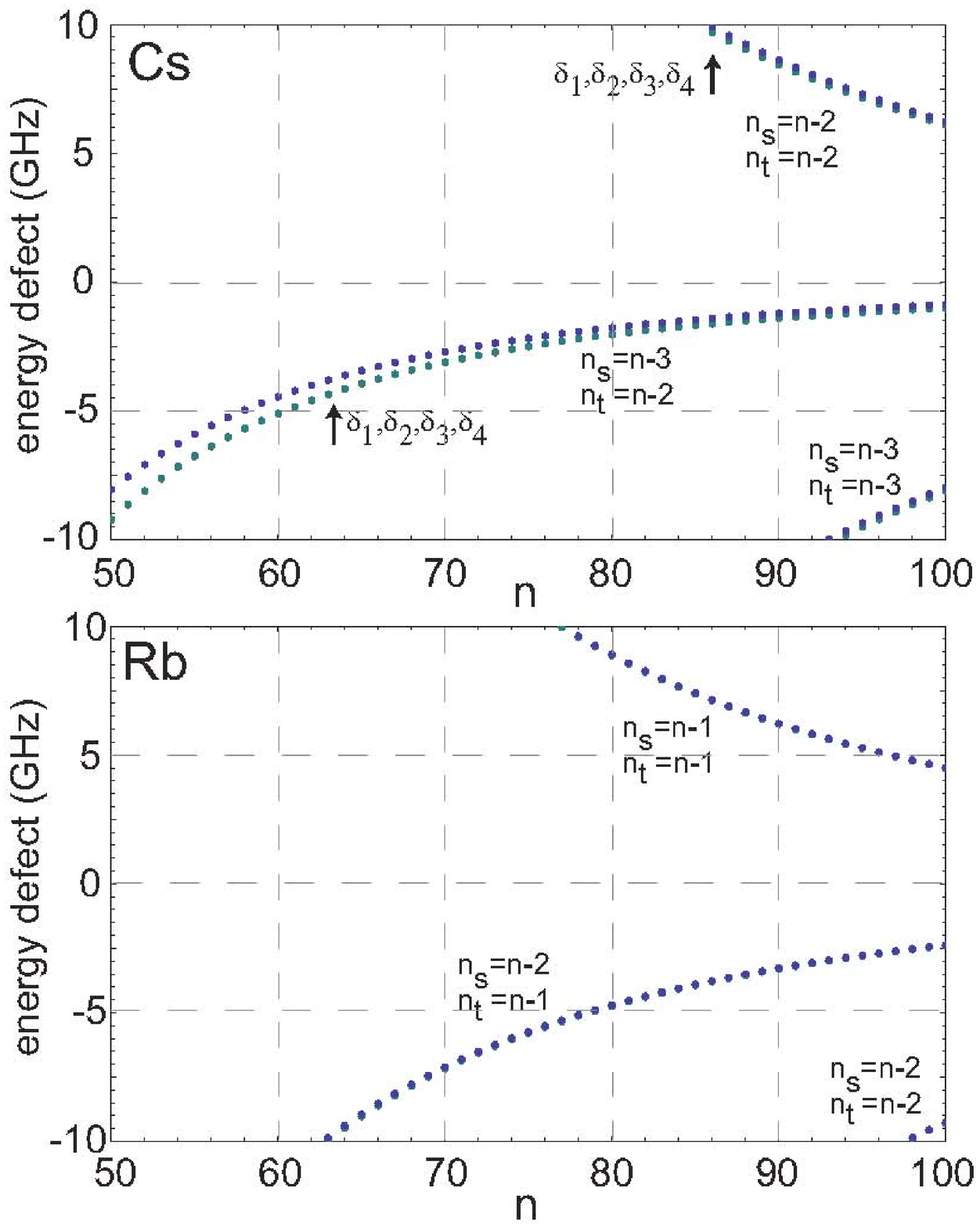}
  \caption{(color online) Energy defects for $nd_{3/2,5/2} + nd_{3/2,5/2}\leftrightarrow n_sf_{5/2,7/2}+ n_tf_{5/2,7/2}$ coupling in Cs and Rb.
 }
\label{fig.dffdefect}
\end{figure}

For Cs the strongest cases are $n_s=n_t=n-3,$ $n_s=n-3,n_t=n-2$, and $n_s=n_t=n-2$ as shown 
in   Fig. \ref{fig.dffdefect}. At $n=70$ we find for $C_6(\delta,n_s,n_t):$
$C_6(\delta_1,67,67)= 15.1 $,
$C_6(\delta_1,67,68)= 188.0 $, and
$C_6(\delta_1,68,68)= -50.5$,
in units of $~\rm GHz~\mu m^6.$ The other channels have similar strengths within about 15\% of the
 given values.

For  Rb the strongest cases are $n_s= n_t=n-2,$ $n_s=n-2,n_t=n-1$, and $n_s= n_t=n-1$.  At $n=70$ we find for $C_6(\delta,n_s,n_t):$
$C_6(\delta_1,68,68)= 7.42 $,
$C_6(\delta_1,68,69)= 77.7 $, and
$C_6(\delta_1,69,69)= -114.$,
in units of $~\rm GHz~\mu m^6.$ The other channels have similar strengths within about 1\% of the given values.

\section{Effective Angular structure  of the F\"orster interaction} \label{sec:angle}

The choice of Rydberg states for blockade experiments is dictated by the strength and angular structure of 
the F\" orster interaction. A three dimensional distribution of atoms includes pairs with arbitrary relative orientations so that laser fields with laboratory fixed polarizations will generally couple to all possible two-atom eigenstates, including those with weak interactions.  Excitation of these F\" orster zero states can be avoided either by choosing Rydberg states with near isotropic interactions, or by using carefully chosen interaction geometries.  We give some representative examples of both approaches in this section.

\begin{figure}[t] 
   \centering
   \includegraphics[width=3.3 in]{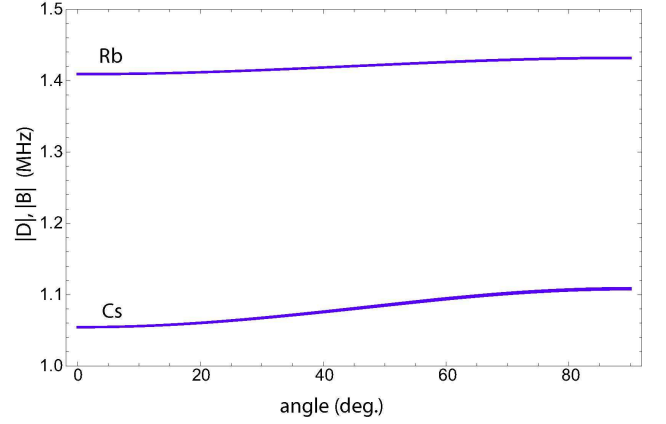} 
   \caption{(color online) Angular dependence of the resonance and blockade shifts  for the $70s_{1/2}$ states in Rb and Cs
at $R=9.2~\mu\rm m.$ 
The $C_6$ parameters for Rb are given in the text and for Cs the values are: 
$C_{61}=712,\, C_{62}=687,$ and $C_{63}=213$,  in units of $\rm GHz\, \mu m^6$. 
}
   \label{fig.spangular}
\end{figure} 

The prime example of a near isotropic interaction is the channel $s_{1/2}+s_{1/2}\rightarrow p+p$ which, as can be seen from Table \ref{evals}, is fully  isotropic provided the fine structure of the $p$ states is ignored. Accounting for fine structure gives the van der Waals Hamiltonian
\bea
H_{\rm vdW}&=&\frac{1}{81 R^{6}}
\left[C_{61}\,{\rm diag}(44,68,36,44)\right.\nonumber\\
&&~~~~~\left.+~C_{62}\,{\rm diag}(28,4,36,28)\right.\nonumber\\
&&~~~~~\left.+~C_{63}\,{\rm diag}(8,32,0,8)  \right].
\eea
 The Hamiltonian has been expressed in the basis 
$(\uparrow\uparrow,$ $\frac{\uparrow\downarrow+\downarrow\uparrow}{\sqrt2},
\frac{\uparrow\downarrow-\downarrow\uparrow}{\sqrt2},
\downarrow\downarrow)$ where the arrows denote the electron spin projections for the two atoms   
and $C_{6j}$ is the coefficient of channel $j$ in Eqs. (\ref{stopchannels}). For the Rb $70s_{1/2}$ state we find $C_{61}=794,\, C_{62}=1125,$ and $C_{63}=427$, and  eigenvalues 
$\{891,862,862,853\}$   in units of ${\rm GHz}\, \mu {\rm m}^6/R^6.$
Excitation of the Rydberg state $70s_{1/2}\ket{\uparrow \uparrow}$ in the laboratory frame  gives the angular
 dependence  shown in Fig. \ref{fig.spangular}. We see that even for Cs which has a relatively large fine structure splitting the interaction strength is close to isotropic. Because of  this the 
resonance shift $\sf D$ and  blockade shift $\sf B$ for each species are almost exactly the same.

  The angular average of the blockade shift in Rb can be approximated by the convenient expression
\begin{equation}
\bshift_{70s}=\mbox{1 MHz}\times \left({9.77 \mbox{ $\mu$}{\rm m}\over R}\right)^{6}
\end{equation}
which shows that a strong blockade is possible for two atoms with separations up to about $10~\mu\rm m.$
In applications to ensembles containing atom pairs with a distribution of $R$ values we  can
calculate the blockade shift analytically using a Gaussian description for the atomic density.
In a spherically symmetric cloud with radial density variance $3\sigma^2$ the probability distribution is $P(r) = {(2\pi \sigma^2)^{-3/2}}e^{-r^2/2\sigma^2}.$ Replacing $\kappa_{\varphi kl}^2/D_\varphi^2$ by an  angular mean $\expect{\kappa^2/D^2}$  the spatially averaged blockade shift is 
\bea
\frac{1}{{\sf B}^2}&\cong&\frac{N}{(N-1)C_6^2}\expect{\frac{\kappa^2}{D^2}}\int_{-\infty}^\infty d{\bf r}\, d{\bf r}'\, P(r) P(r')|{\bf r}-{\bf r}'|^{12}\nonumber\\
&=&\frac{N}{(N-1)}\expect{\frac{\kappa^2}{D^2}}\frac{(3.785\, \sigma)^{12}}{C_6^2},
\eea
where $N$ is the number of atoms and we have used $8648640^{1/12}\cong 3.785.$
For the $s_{1/2}+s_{1/2}\rightarrow p+p$ channel the interaction is isotropic and $\kappa^2/D^2=9/16$.
Note that the strong weighting of the integrand towards large $|{\bf r}-{\bf r}'|$ implies that the assumption of the van der Waals form for all molecular separations holds approximately even though at 
small separation the interactions may transition into the resonant \forster\ regime.

A second spatial distribution of interest is a quasi one-dimensional ensemble in an optical trap created by  tightly focused laser beams. 
 If $T_{\rm rel},$ the 
atomic temperature relative  to the peak depth of the confining potential, is small we can use a quadratic approximation to the potential about its minimum which leads to a Gaussian distributed density.
To be specific consider a far off resonance trap  (FORT) created by focusing a single Gaussian beam of wavelength $\lambda$ to a 
waist $w$ ($1/e^2$ intensity radius). When $w$ is at least several times larger than $\lambda$ the 
trap provides a quasi one-dimensional distribution with probability density $P(z)={(2\pi \sigma^2 )^{-1/2}}e^{-z^2/2\sigma^2}$ where 
$\sigma=\frac{\pi w^2}{2^{1/2}\lambda}T_{\rm rel}^{1/2}$. FORT traps were used in several recent experiments with highly localized atomic clouds\cite{Sebby05,Yavuz06,Johnson07}. 
The blockade shift  then takes the form 
\bea
\frac{1}{{\sf B}^2}&=&\frac{N\left(\kappa^2/D^2\right)_\theta}{(N-1)C_6^2}\int_{-\infty}^\infty dz\, dz' P(z) P(z')(z-z')^{12}\nonumber\\
&=&\frac{N}{(N-1)}\left(\frac{\kappa^2}{D^2}\right)_\theta\frac{(3.0567\, \sigma)^{12}}{C_6^2},
\eea
where $\left(\kappa^2/D^2\right)_\theta$ gives the interaction strength when the trap is tilted by an angle $\theta$ with respect to the quantization axis $\hat z$ of the light,   and we have written   $665280^{1/12}\cong 3.0567.$ 
A single beam FORT with $T_{\rm rel}=2.5, w=2.5~\mu\rm m,$ and $\lambda=1.03~\mu\rm m$ which is close to the parameters of our recent experiment\cite{Johnson07} gives $\sigma=3.0~\mu\rm m$ so $3.0567\, \sigma=9.2~\mu\rm m$ and the averaged interaction strength for 70s is that shown in Fig. \ref{fig.spangular}.

\begin{figure}[t] 
   \centering
   \includegraphics[width=3.3 in]{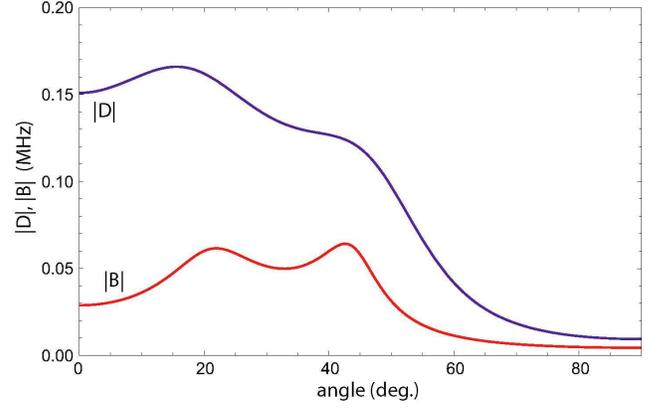} 
   \caption{(color online) Angular dependence of the resonance and blockade shifts  for the $43d_{5/2}$ state in Rb assuming a one-dimensional atom cloud with  $\sigma=3.0~\mu\rm m.$  
}
   \label{fig.43dangular}
\end{figure} 

In contrast to the $s$ states the other channels in Table \ref{evals} exhibit strong angular effects.
A case of particular  interest is 43d$_{5/2}$+ 43d$_{5/2}\rightarrow$ 45p$_{3/2}$ +41f$_{j}$, which is within a few MHz of being \forster\ resonant for Rb.  The \forster\ defects are -6.0, -8.3 MHz for $j=5/2,7/2$, which give $C_6$ coefficients of $C_{61}=391,$ $C_{62}=539~ {\rm GHz}\, \mu {\rm m}^6$ with $C_{6j}$ referring to channel $j$ in Eqs. (\ref{pdfchannels}).  The effective $C_6$, averaging over the degeneracy of the two channels, is $C_6=\frac{4}{7}C_{61}+\frac{3}{7}C_{62}=454~ {\rm GHz}\, \mu {\rm m}^6$. If we were to ignore angular effects we would naively expect a strong interaction  of order $454/860=0.53$ times that shown in Fig. \ref{fig.spangular}, but at a much smaller value of $n$ which relaxes the laser power requirements for fast excitation.  

However, as mentioned above, the 36 $D_\varphi$ coefficients for the $d_{5/2}+d_{5/2}\rightarrow p_{3/2}+f$ channel cover a huge range, with two $M=0$ states being extremely small. To be explicit assume we start in the $^{87}$Rb $f=2, m_f=0$ hyperfine ground state and use $\pi$-polarized excitation light which couples to the  Rydberg state (expressed in a basis aligned with the light polarization)
\begin{eqnarray}
\ket{\gamma_{k}\gamma_{l}}&=&{\half1}\left(\ket{\half{1}\half{1};\half{-1}\half{-1}}+\ket{\half{1}\half{-1};\half{-1}\half{1}}\right.\nonumber \\
&&+\left.\ket{\half{-1}\half{1};\half{1}\half{-1}}+\ket{\half{-1}\half{-1};\half{1}\half{1}}\right)
\label{twoatomket}
\end{eqnarray}
where  the second two numbers in each  ket are the magnetic quantum numbers for the nuclear spin.  The latter are conserved in the excited state (assuming the hyperfine interaction can be neglected there).  We must therefore average $1/\bshift^{2}$ for each of the four terms to calculate the angular dependence shown in Fig. \ref{fig.43dangular}. The small value of $\sf B$ for angles between the light polarization and the molecular axis near $90 ~\rm deg.$ render this interaction a  poor choice for blockade in a spherical ensemble. 
We see that even at the optimum angle the blockade shift is only  $65~\rm kHz$ which is about 20 times smaller than in Fig. \ref{fig.spangular}. Furthermore, the resonance shift $\sf D$ is  larger than the blockade shift, and has a somewhat different angular structure.

\begin{figure}[t] 
   \centering
   \includegraphics[width=3.3 in]{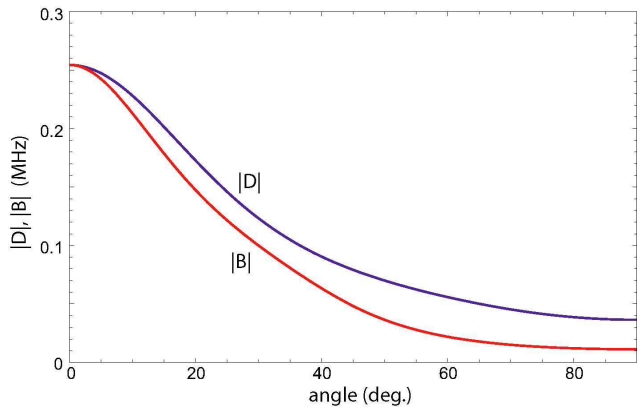} 
   \caption{(color online) Angular dependence of the resonance and blockade shifts  for the $43d_{5/2}$ state in Rb for excitation of the $\ket{\half5,\half5}$ Rydberg state and assuming a one-dimensional atom cloud with  $\sigma=3.0~\mu\rm m.$  
}
   \label{fig.43dstretched}
\end{figure} 

It is worth noting that it is possible to choose interaction geometries which largely avoid the
small $D_\varphi$ coefficients of $0.003, 0.002$ for this channel.  
 The state with the smallest $D_\varphi$ is interchange symmetric  (the eigenvector can be found in the accompanying EPAPS material) 
\begin{eqnarray}
\ket{\psi_{F}}=\text{0.67} \ket{-\frac{5}{2},\frac{5}{2}}
+\text{0.20}\ket{-\frac{3}{2},\frac{3}{2}}
+\text{0.08}\ket{-\frac{1}{2},\frac{1}{2}}&&\nonumber\\
+\text{0.08}\ket{\frac{1}{2},-\frac{1}{2}}
+\text{0.20}\ket{\frac{3}{2},-\frac{3}{2}}
+\text{0.67} \ket{\frac{5}{2},-\frac{5}{2}}&&
\label{psif}
\end{eqnarray}
where the kets are given in the form $\ket{m_{k},m_{l}}$.  The state with the second smallest $D_\varphi$ 
is very similar to the above but is interchange antisymmetric. Coupling to $\ket{\psi_F}$ can be strongly suppressed by using $\sigma^+$ excitation light with a one-dimensional trap aligned along $\hat z$, and Zeeman selecting a target $m_j$ level to excite the two-atom state $\ket{\frac{5}{2},\frac{5}{2}}.$ The angular distribution for this state is shown in Fig. \ref{fig.43dstretched}. We see that for a trap aligned parallel to the quantization axis the blockade strength is $0.25~\rm MHz$, which is more than 8 times larger than when exciting the state given in  Eq. (\ref{twoatomket}). Finally, we note that a similar trick can be used to render the $58d_{3/2}+58d_{3/2}\rightarrow 60p_{1/2}+56f_{5/2}$ resonance discussed above usable for blockade in one-dimensional geometries. The blockade shift obtained at $\theta=0$ by exciting $\ket{\frac{3}{2},\frac{3}{2}}$  is a very large  $|B|=2.9~\rm MHz$ in a trap with $\sigma=3~\mu\rm m.$

\section{Conclusions}

In this paper we have considered in detail the effects of Zeeman degeneracy on  blockade experiments relying on van der Waals interactions to allow only single-atom excitations.  The figure of merit for blockade is sensitive primarily to the weakest interactions between the various degenerate Rydberg states.
For many convenient Rydberg states, the degeneracies result in particular linear combinations of Zeeman sublevels having zero or nearly zero van der Waals interactions.  This problem can sometimes be avoided using special geometries and choices of light polarization, but care must be taken.

We have catalogued the long-range potential curves for a large number of angular momentum channels likely to be of interest to blockade experiments, with sufficient information to allow researchers to quantitatively evaluate the van der Waals interactions for a wide range of experimental situations.

All blockade experiments reported to date have used samples whose spatial extents are substantially larger than the range of the van der Waals interactions.  In these situations, dipole-dipole or van der Waals interactions can play a dominant role in suppressing Rydberg excitation under conditions where quantum blockade would not be possible.

\begin{acknowledgments}This research was supported by the National Science Foundation and ARO-IARPA.  We appreciate helpful conversations with Deniz Yavuz and other members of the Wisconsin Quantum Computing group.
\end{acknowledgments}

\appendix\section{Quadrupole-Quadrupole Interaction} \label{quadquad}

 There is also a quadrupole-quadrupole (Q-Q) interaction that contributes a $R^{-5}$ term in the long-range potential \cite{Zhang07}:
 \begin{equation}
V_{QQ}={\sqrt{70}\over R^{5}}\tp{\bm Q_{a}}{\bm Q_{b}}{40}={\sqrt{70}\over R^{5}}\sum_{p}\CG{2p}{2\bar p}{40}Q_{ap}Q_{b\bar p}
\end{equation}
where the atomic quadrupole moment operator is $Q_{p}=er^{2}\sqrt{4\pi/5}Y_{2p}$.
As with the van der Waals interaction, the quadrupole-quadrupole interaction causes transitions between different Zeeman levels.  

The following argument shows that for most cases the quadrupole-quadrupole interaction will be considerably smaller than the van der Waals.  For the Q-Q interaction to dominate over the van der Waals interaction, we need
\begin{equation}
{e^{2}\expect{r^{2}}^{2}\over R^{5} }\gg{e^{4}\expect{r}^{4}\over \delta  R^{6} }
\end{equation}
The matrix element factors are roughly equal, so we find
\begin{equation}
R\gg {e^{2}\over \delta}\sim 350\mbox{ $\mu$m}
\end{equation} for a 1 GHz (or smaller) value of $\delta$ that is usual for Rydberg states. Thus the Q-Q interaction should be negligible.

\section{Dipole-Dipole Interaction in the Coupled Basis}\label{coupled}

Instead of the basis $\ket{j_{a}m_{a}j_{b}m_{b}}$ we could equally well do the calculations in the coupled basis $\ket{j_{a}j_{b}JM}$.  Using multipole tensors and recoupling algebra, a simple relation for the matrix elements of the dipole-dipole interaction can be obtained.

The multipole tensors are defined using the tensor product formalism of \Varsh{3.1.7}:
\begin{equation}
T_{JM}^{j'j}=\tp{\ket{j'}}{\bbra{j}}{JM}=\sum_{m}C_{j'mjM-m}^{JM}\ket{jm}\bbra{j'M-m}
\end{equation}
where the time reversed bra is defined as
\begin{equation}
\bbra{jm}=(-1)^{j+m}\bra{j\bar m}
\end{equation}
The spherical component of the position operator $\bf r$ of an electron can, for example, be written in terms of $T_{1}$:
\begin{equation}
r_{p}=\sum_{j_{s},j}{\bra{j_{s}}|r|\ket{j}\over \sqrt{3}}T_{1p}^{j_{s}j}
\end{equation}
as can be verified by taking matrix elements of both sides of the equation.

The dipole-dipole interaction  is proportional to the spherical tensor
\begin{eqnarray}
\tp{\bm a}{\bm b}{20}&=&{\bra{j_{s}}|r|\ket{j}\bra{j_{t}}|r|\ket{j}\over {3}}\tp{T_{1}^{j_{s}j}}{T_{1}^{j_{t}j}}{20} \nonumber\\
&=&{\bra{j_{s}}|r|\ket{j}\bra{j_{t}}|r|\ket{j}\over {3}}\nonumber\\
&&\times \tp{\tp{\ket{j_{s}}}{\bbra{j}}{1}}{\tp{\ket{j_{t}}}{\bbra{j}}{1}}{20}
\end{eqnarray} where we are assuming that in the initial states the electrons have the same angular momentum $j$, and the coupling of interest is isolated to a single state where the electrons on atoms $a$ and $b$ have angular momenta $j_{s}$ and $j_{t}$.

We can now use recoupling algebra to rewrite this in terms of the coupled states $\tp{\ket{j_{s}}}{\ket{j_{t}}}{KM}=\ket{j_{s}j_{t}KM}$ and $\tp{\ket{j}}{\ket{j}}{JM}=\ket{jjJM}$, using \Varsh{3.3.2 (11)}:
\begin{eqnarray}
{\tp{\bm a}{\bm b}{20}\over {\bra{j_{s}}|r|\ket{j}\bra{j_{t}}|r|\ket{j}}}&=&\sum_{KJ}\sqrt{[J][K]} \ninej{j_{s}}{j}{1}{j_{t}}{j}{1}{K}{J}{2}\tp{\ket{K}}{\bbra{J}}{20}
\nonumber\\
&=&\sum_{KJ}\sqrt{[J][K]} \ninej{j_{s}}{j}{1}{j_{t}}{j}{1}{K}{J}{2}T_{20}^{KJ}
\end{eqnarray}
where $[J]=\sqrt{2J+1}$.

In this basis, it is often found that the van der Waals eigenstates are heavily weighted with a single value of $J$, suggesting that in many cases $J$ is an approximately good quantum number.

\input{vanderwaals2.bbl}

\end{document}

%% file: vanderwaals2.bbl
\newcommand{\noopsort}[1]{} \newcommand{\printfirst}[2]{#1}
  \newcommand{\singleletter}[1]{#1} \newcommand{\switchargs}[2]{#2#1}